\documentclass[a4paper,nofootinbib]{revtex4-1}
\usepackage[utf8]{inputenc}
\usepackage{graphicx,bm,color}
\usepackage{slashed,verbatim}
\usepackage{amssymb,graphicx,epstopdf}
\usepackage[colorlinks=true,linktocpage=true,linkcolor=blue,citecolor=blue]{hyperref}
\usepackage[margin=0.9in]{geometry}
\usepackage{ulem}
\def\be {\begin{equation}}
\def\ee {\end{equation}}
\def\bea {\begin{eqnarray}}
\def\eea {\end{eqnarray}}
\def\bc {\begin{center}}
\def\ec {\end{center}}
\def\nn {\nonumber}

\begin{document}

\title{Electromagnetic spectral properties and Debye screening of a 
strongly magnetized 
hot medium}
\author{Aritra Bandyopadhyay}
\author{Chowdhury Aminul Islam}
\author{Munshi G. Mustafa}
\affiliation{
Theory Division, Saha Institute of Nuclear Physics, HBNI,
Kolkata 700064, India.
}
\email{aritra.bandyopadhyay@saha.ac.in}
\email{chowdhury.aminulislam@saha.ac.in}
\email{munshigolam.mustafa@saha.ac.in}
\date{\today}

\begin{abstract}
We have evaluated the electromagnetic spectral function and its spectral properties 
by computing the one-loop photon polarization tensor involving quarks in the loop, 
particularly in a strong field approximation compared 
to the thermal scale. When the magnetic scale is higher than the thermal scale
the lowest Landau level (LLL) becomes effectively (1+1) dimensional strongly correlated 
system that provides a kinematical threshold based on the quark mass scale. Beyond this 
threshold the photon strikes 
the LLL and the spectral strength starts with a high value due to the dimensional 
reduction and then falls off with increase of the photon energy due to LLL dynamics in a 
strong field approximation. We have obtained analytically the dilepton production rates
from LLL considering the lepton pair remains  unaffected by the magnetic field when 
produced at the edge of a hot magnetized medium  or affected by the magnetic
field if produced inside a hot magnetized medium. For the later case the production rate 
is of ${\cal O}[|eB|^2]$ along with an additional kinematical threshold due to lepton 
mass than the former one.
We have also investigated the 
electromagnetic screening  by computing the Debye screening mass and it
depends distinctively on three different scales (mass of the quasiquark, temperature and 
the 
magnetic field strength) of a hot magnetized system. The mass dependence of the Debye 
screening supports the occurrence of a magnetic catalysis effect in the strong field 
approximation.

\end{abstract}
\maketitle


\section{Introduction}
\label{intro}
Ongoing relativistic heavy ion collisions provide enough indications of the formation  of 
the deconfined state of hadronic 
matter called Quark Gluon Plasma (QGP) and the nuclear matter under extreme conditions has 
been a subject of scrutiny. 
Recent studies \cite{Shovkovy,Elia,Fukushima,Mueller,Miransky} have revealed a captivating 
nature of non-central heavy ion collisions (HIC). 
It indicated 
that in such collisions, a very strong anisotropic magnetic field is generated in the direction perpendicular 
to the reaction plane, due to the relative motion of the ions themselves. The initial magnitude of this magnetic field can be very high 
($eB\approx m_\pi^2$ at RHIC and $eB\approx 10m_\pi^2$ at LHC) at the time of the collision and then it decreases very fast, being 
inversely proportional to the square of time \cite{bzdak,McLerran}\footnote{However for a 
different point of view, see~\cite{Tuchin:2013prc,Tuchin:2013prc2,Tuchin:2013ahep}, where 
the time 
dependence of magnetic field is shown to be adiabatic due to the high conductivity of 
the medium.}. 

The presence of an external anisotropic field in the medium subsequently requires 
modification of the present theoretical 
tools that can be applied appropriately to investigate various properties of QGP. 
An intense research activity is underway  to study the properties of strongly interacting 
matter in presence of an external magnetic field: resulting the emergence 
of several novel phenomena, \textit{e.g}, chiral magnetic effect~\cite{cme1,cme2,cme3}, 
finite 
temperature magnetic catalysis~\cite{mcat1,mcat2,mcat3} and inverse magnetic 
catalysis~\cite{Bali,Bornyakov,mueller,Ayala:2014iba,Ayala:2014gwa,Ayala:2016sln,
Ayala:2015bgv}; 
chiral- and 
color-symmetry broken/restoration 
phase~\cite{fayazbakhsh1,fayazbakhsh2,jens1};  
thermodynamic 
properties~\cite{mike1,jens_rmp}, refractive indices and decay 
constant~\cite{fayazbakhsh3,fayazbakhsh4}
of mesons in hot magnetised medium; soft photon production from conformal 
anomaly~\cite{basar,Ayala:2016lvs} in HICs; modification of disperson properties in a 
magnetised hot QED medium~\cite{sadooghi1}; syncroton 
radiation~\cite{Tuchin:2013prc2}, dilepton production from a hot magnetized QCD 
plasma~\cite{Tuchin:2013prc,Tuchin:2013prc2,Tuchin:2013ahep,sadooghi} and in strongly
coupled plasma in a strong magnetic field~\cite{mamo}.
Also experimental 
evidences of photon anisotropy, 
provided by the PHENIX
Collaboration~\cite{phenix}, has posed a challenge for existing theoretical models. 
Subsequently some theoretical explanations are made by assuming the presence of a large 
anisotropic magnetic field in heavy ion collisions~\cite{basar}.
This suggests that  there is clearly an increasing demand to study the effects of
intense background magnetic fields on various aspects and observables
of non-central heavy-ion collisions.

We know that the energy levels (orbital) of a moving charged particle  in 
presence of a magnetic field get discretized, which are known as the Landau Levels (LL). 
One fascinating prospect of having a 
very strong background magnetic field is that only the Lowest Landau Level (LLL), whose 
energy is independent of the strength of the magnetic field, remains active in that 
situation.
That is why, the LLL dynamics becomes solely important in the strong magnetic 
field approximation and the higher order contributions, \textit{i.e}, the radiative 
corrections play 
a significant role in this context, as it is the only way to get the $B$ dependence in 
the LLL energy.

One primary ingredients of the theoretical tools for studying various properties of QGP is 
the $n$-point correlation function, which eventually determines the laws 
of propagation and the thermodynamic potential. Among them the electromagnetic 
correlation function is of particular interest because it is related to various physical 
quantities associated with the deconfined state of matter.
As for example the production rate
of real and virtual photons (dilepton pairs therefrom), which leave the fireball with 
minimum interaction.
These electromagnetic probes are produced in every stages of the HICs. The 
dilepton spectra is a space-time integrated observable which has 
contributions coming from  various stages of 
the collisions. Even though the dilepton may carry almost undistorted 
information of the stages they are produced, it would be very difficult to disentangle 
the contribution from different stages.

Processes like cyclotron emission which are usually abandoned in vacuum become active in 
presence of an external magnetic field~\cite{rojas}. These processes affect the photon 
propagation and thus the spectral function. The spectral function or the spectral 
discontinuity of the electromagnetic correlator is directly related to the production rate 
of dileptons and photons. In vacuum, a full description of polarization tensor in presence 
of an external magnetic field have already been studied~\cite{hattori,chao,Tsai1,Tsai2}. 
In this article we, first, would like to obtain  the spectral representation of 
the electromagnetic correlation function in presence of a strong  background 
magnetic field at finite temperature. As a spectral property we then calculate the 
dilepton rate which is of immense importance especially in the scenario of non-cnetral
heavy-ion collisions. At this point we note that 
the dilepton production rate 
under extreme magnetic fields have  been addressed  earlier by
Tuchin~\cite{Tuchin:2013prc,Tuchin:2013prc2,Tuchin:2013ahep} in a more
phenomenological way. In order to estimate the dilepton production with
logarithmic accuracy~\cite{Tuchin:2013prc,Tuchin:2013ahep}, a semi-classical 
Weisz\"acker-Williams method~\cite{jackson}
was employed to obtain
the dilepton production rate by a hard quark as a convolution of 
the real photon decay rate with the flux of equivalent photons emitted by a fast
quark. In this calculation it was approximated that the virtuality of 
photon has neglibible effect on photon emission and on dilepton production. Recently, 
Sadooghi and Taghinavaz~\cite{sadooghi}
have  analyzed in details the dilepton production rate 
for magnetized hot and dense medium in a formal field theoretic approach 
using Ritus eigenfunction method~\cite{ritus}.
In this article we use such formal field theoretic approach along with Schwinger 
method~\cite{schwinger} to obtain the 
electromagnetic 
spectral function and the dilepton rate in strong field approximation and compare our 
results with those of Ref.~\cite{sadooghi}.
In addition we also discuss another interesting topic, namely the Debye screening,  which 
could reveal some of the 
intriguing properties of the medium in presence of strong magnetic field.

The paper is organized as follows: in sec.~\ref{setup}  we briefly review the setup,
within the Schwinger formalism~\cite{schwinger}, required to compute the photon 
polarization tensor
in presence of a very strong background magnetic field along the $z$ direction. In 
sec.~\ref{sfa}  we briefly discuss the vacuum spectral function and then obtain the 
in-medium photon polarization tensor and its spectral 
representation in strong field approximation. In sec.~\ref{dil} we discuss
how the dilepton rate for LLL approximation would be modified  and
calculate the analytic expression for the dilepton production rate for various 
scenarios~\cite{Tuchin:2013prc} in the strong magnetic field approximation. 
A closer view of the Debye screening in a strongly magnetized hot medium  is 
taken up in sec.~\ref{ds} before concluding in sec.~\ref{conclu}.


\section{Setup}
\label{setup}

In presence of a constant magnetic field pointing towards the $z$ direction 
($\vec{B}=B\hat{z}$), we first describe the charged fermion propagator.  In 
coordinate space it can be 
expressed~\cite{schwinger} as
\bea
S_m(x,x^\prime) = e^{\Phi(x,x^\prime)}\int\frac{d^4k}{(2\pi)^4}e^{-ik(x-x^\prime)}S_m(k) , 
\eea
where $\Phi(x,x^\prime)$ is called the phase factor, which generally drops out in gauge 
invariant correlation functions and the exact  
form of $\Phi(x,x^\prime)$ is not important in our problem. In momentum space the 
Schwinger propagator $S_m(k)$ can be written~\cite{schwinger} as an 
integral over proper time $s$, i.e.,
\bea
i S_m(k) = \int\limits_0^\infty ds \exp\left[i s \left(k_\shortparallel^2-m_f^2-\frac{k_\perp^2}{q_f Bs}\tan(q_f Bs)\right)\right] \nn\\
\times \left[\left(\slashed{k}_\shortparallel+m_f\right)\left(1+\gamma_1\gamma_2\tan(q_f B s)\right)-\slashed{k}_\perp\left(1+\tan^2(q_f Bs)\right)\right].
\label{schwinger_propertime}
\eea
Here, $m_f$ and $q_f$ are the mass~\footnote{ Even if there is a dynamical mass 
generation in the system, one needs to take appropriate modification. However,  
the fermion mass is generically represented by $m_f$ in this calculation.} and 
\textit{absolute charge} of the fermion of flavor $f$, respectively. Below we outline the 
notation 
we have used in~(\ref{schwinger_propertime}) and  are going to follow throughout as
\bea
&&a^\mu = a_\shortparallel^\mu + a_\perp^\mu;~~ a_\shortparallel^\mu = (a^0,0,0,a^3) ;~~  a_\perp^\mu = (0,a^1,a^2,0),\nn\\
&&g^{\mu\nu} = g_\shortparallel^{\mu\nu} + g_\perp^{\mu\nu};~~ g_\shortparallel^{\mu\nu}= \textsf{diag}(1,0,0,-1);~~ g_\perp^{\mu\nu} = \textsf{diag}(0,-1,-1,0),\nn\\
&&(a\cdot b) = (a\cdot b)_\shortparallel - (a\cdot b)_\perp;~~ (a\cdot b)_\shortparallel = a^0b^0-a^3b^3;~~ (a\cdot b)_\perp = (a^1b^1+a^2b^2),\nn
\eea
where $\shortparallel$ and $\perp$ are, respectively, the parallel and perpendicular 
components, which are now separated out in momentum space propagator.
After performing the proper time integration~\cite{gusynin}, the fermion propagator in (\ref{schwinger_propertime}) can be represented as 
sum over discrete energy spectrum of the fermion 
\bea
i S_m(k) = i e^{-\frac{k_\perp^2}{q_fB}} \sum_{n=0}^{\infty} \frac{(-1)^nD_n(q_fB, k)}{k_\shortparallel^2-m_f^2-2nq_fB},
\label{decomposed_propagator}
\eea
with Landau levels $n=0,\, 1,\, 2, \cdots$ and  
\bea
D_n(q_fB,k) &=& (\slashed{k}_\shortparallel+m_f)\Bigl((1-i\gamma^1\gamma^2)L_n\left(\frac{2k_\perp^2}{q_fB}\right)
-(1+i\gamma^1\gamma^2)L_{n-1}\left(\frac{2k_\perp^2}{q_fB}\right)\Bigr)
- 4\slashed{k}_\perp L_{n-1}^1\left(\frac{2k_\perp^2}{q_fB}\right),
\label{d_n}
\eea
where $L_n^\alpha (x)$ is the generalized Laguerre polynomial written as
\bea
(1-z)^{-(\alpha+1)}\exp\left(\frac{xz}{z-1}\right) = \sum_{n=0}^{\infty} L_n^\alpha(x) z^n.
\eea
The energy level of charged fermions in presence of magnetic field follows from the pole of the propagator
in (\ref{decomposed_propagator}) as
\bea
&& k_\shortparallel^2-m_f^2-2nq_fB= k_0^2-k_3^2-m_f^2-2nq_fB= 0  \nonumber \\
&& \Longrightarrow \, E_n=k_0 =\sqrt{k_3^2+m_f^2+2nq_fB} . \label{discrete_ll}
\eea
As seen that the energy along the direction of the magnetic field $(0,\, 0,\, B)$ is continuous but  discretized 
along the transverse direction of the field. These discretized energy levels are so called Landau levels, 
which are degenerate for each value of $k_3$.  
These Landau levels can affect  the quantum fluctuations of the charged fermions in the Dirac sea 
at $T=0$ and thermal fluctuations at $T\ne 0$, both of which arise as a response 
to the polarization of the electromagnetic field. These fluctuations are usually 
related to the electromagnetic polarization tensor or the self energy of 
photon, which in one loop level is  expressed as
\bea
\Pi_{\mu\nu}(p) = -i\sum_{f} q_f^2\int \frac{d^4k}{(2\pi)^4} 
\textsf{Tr}_{c}\left[\gamma_\mu S_m(k) 
\gamma_\nu S_m(q)\right] , \label{pola}
\eea
where $p$ is the external momentum, $k$ and $q=k-p$  are the loop 
momenta. $\textsf{Tr}_{c}$ represents both  color and Dirac traces whereas 
the $\sum_{f}$ is over flavor  because   
we have considered a two-flavor system ($N_f=2$)  of 
equal current quark mass ($m_f=m_u=m_d=5$ MeV if not said otherwise).

The two point current-current correlator $C_{\mu\nu}(p)$ is related to
photon self-energy as 
\bea
q_f^2C_{\mu\nu} (p) &=&  \Pi_{\mu \nu} (p), \label{corr_func}
\eea
with $q_f$ is the electric charge of a given quark flavor $f$.
The electromagnetic spectral representation is extracted from the imaginary part of 
the correlation function $C_\mu^\mu(p)$ as
\bea
\rho(p) &=& \frac{1}{\pi} \mathcal{I}m~C^\mu_\mu(p)=\frac{1}{\pi} 
\mathcal{I}m~\Pi^\mu_\mu(p)/q_f^2.  \label{spec_func}
\eea


\section{Electromagnetic spectral function and its properties in presence of strong background magnetic field}
\label{sfa}
In this section we will mainly 
investigate the nature of the in-medium electromagnetic 
spectral function in presence of a very 
strong but constant magnetic field strength ($q_fB~\gg~T^2$), which could be 
relevant for initial stages
of a non-central heavy-ion collisions, as a high  intensity magnetic field is 
believed to be produced there.

\begin{figure}
\begin{center}
\includegraphics[scale=1.2]{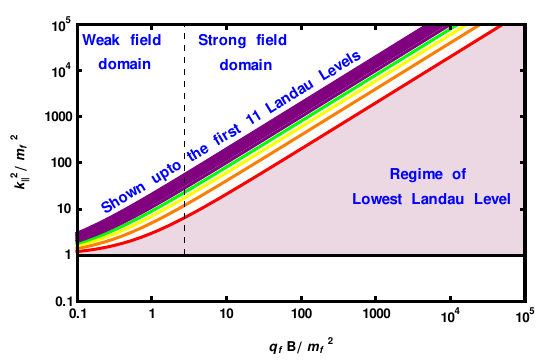}
\end{center}
\caption{Thresholds corresponding to a few Landau Levels are displayed as a function of $q_fB/m_f^2$.
This threshold plot is obtained by solving $\left(\omega^2-4m_f^2-8nq_fB\right)=0$ with 
zero photon momentum following energy conservation in a background magnetic field in 
general.
Also the regime of the  LLL at strong magnetic field approximation is 
shown by the shaded area.}
\label{landau_levels}
\end{figure}

When the external magnetic field is very strong~\cite{calucci}, $q_fB\rightarrow \infty$, 
it pushes all the Landau levels ($n\ge 1$) to infinity
compared to the Lowest Landau Level (LLL) with $n=0$ (See Fig.\ref{landau_levels}). For LLL approximation in the strong field limit the 
fermion propagator in (\ref{decomposed_propagator}) reduces to a simplified form as

\bea
iS_{ms}(k)=ie^{-{k_\perp^2}/{q_fB}}~~\frac{\slashed{k}_\shortparallel+m_f}{
k_\shortparallel^2-m_f^2}(1-i\gamma_1\gamma_2),
\label{prop_sfa}
\eea
where $k$ is four momentum and we have used the properties of generalized Laguerre 
polynomial, $L_n\equiv L_n^0$ and $L_{-1}^\alpha = 0$. One could  also get to 
(\ref{prop_sfa}) directly from (\ref{schwinger_propertime}) by putting $q_fB\rightarrow 
\infty$. The appearance of the projection operator $(1-i\gamma_1\gamma_2)$ in 
(\ref{prop_sfa}) indicates that the spin of the fermions in LLL are aligned along the 
field 
direction~\cite{Shovkovy,gusynin}.
As $k_\perp^2 << q_fB$,  one can see from (\ref{prop_sfa}) that an effective dimensional 
reduction from (3+1) to (1+1) takes place in the strong field limit. 

\begin{figure}
\begin{center}
\includegraphics[scale=0.7]{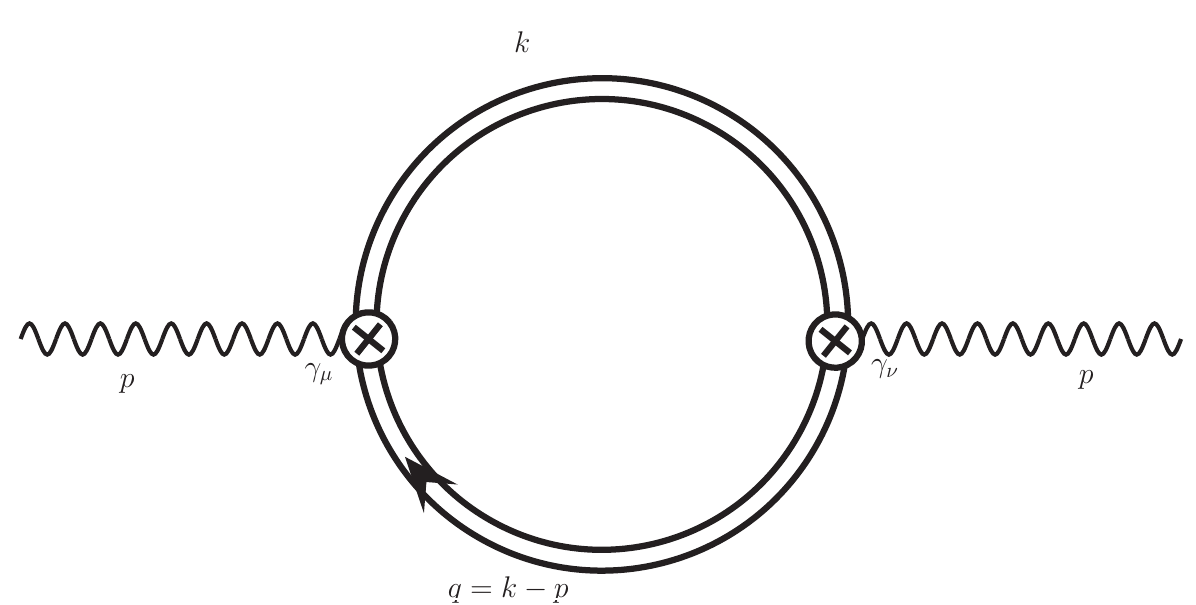}
\end{center}
\caption{Photon polarization tensor in the limit of strong field approximation.}
\label{sfa_self_energy}
\end{figure}

As a consequence the motion of the charged particle is restricted in the direction perpendicular
to the magnetic field  but can move along the field direction in LLL.
This effective dimensional reduction
also plays an important role in catalyzing the spontaneous chiral symmetry 
breaking~\cite{Shovkovy,gusynin} 
since the fermion pairing takes place in LLL, which enhances the generation of 
fermionic mass through the chiral condensate in strong field limit at $T=0$.
The pairing dynamics is essentially (1+1) dimensional where the fermion pairs fluctuate in the direction 
of magnetic field. It is also interesting to see how these fermionic pairs respond to 
the electromagnetic fields. The fluctuation of fermion pairs
in LLL as shown in Fig.~\ref{sfa_self_energy} is a response to the polarization of the 
electromagnetic field and would reveal various properties of the system in 
presence of 
magnetic field. Also the response to the electromagnetic field at $T\ne 0$ 
due to the thermal fluctuation of charged fermion pairs in LLL would also be very 
relevant for the initial 
stages of the noncentral heavy-ion collisions where the intensity of the generated magnetic field is very high. 

Now in one-loop photon polarization in Fig.~\ref{sfa_self_energy} the effective fermionic 
propagator in strong field approximation
is represented by a doubled line and the electromagnetic vertex remains 
unchanged~\footnote{This is not very apparent from the momentum space effective 
propagator in (\ref{prop_sfa}) because of the presence of the projection
operator. In 
Ref.\cite{ferrer} the Ward-Takahasi identity in  LLL 
for fermion-antifermion-gauge boson in massless QED in presence of constant magnetic 
field was shown to be satisfied by considering the effective fermion propagator, bare 
vertex and free gauge boson propagator in ladder approximation through Dyson-Schwinger
approach  in a representation where the fermion mass operator is diagonal in momentum 
space.} and denoted  by a crossed circle.
As mentioned earlier that the spin of the fermions in LLL are aligned in the direction of 
the magnetic field because of the projection operator in (\ref{prop_sfa}). In QED like 
vertex with two fermions from LLL make the photon spin equals to zero 
in the field direction~\cite{gusynin} and there is no polarization in the transverse 
direction.  Thus the longitudinal components ({\it i.e}, (0,3)-components) of QED vertex 
would  only be relevant. 

Now in the strong field limit the self-energy in (\ref{pola}) can be computed as
\bea
\Pi_{\mu\nu}(p)\Big\vert_{sfa} &=& -i\sum_{f}q_f^2\int\frac{d^4k}{(2\pi)^4}\textsf{Tr}_c\left[\gamma_\mu S_{ms}(k)\gamma_\nu S_{ms}(q)\right]\nn\\
&=& -iN_c\sum_{f}q_f^2 \int\frac{d^2k_\perp}{(2\pi)^2} \exp\left(\frac{-k_\perp^2-q_\perp^2}{q_fB}\right)\nn\\
&&\times \int\frac{d^2k_\shortparallel}{(2\pi)^2} \textsf{Tr} \left[\gamma_\mu \frac{\slashed{k}_\shortparallel+m_f}
{k_\shortparallel^2-m_f^2}(1-i\gamma_1\gamma_2)\gamma_\nu \frac{\slashed{q}_\shortparallel+m_f}{q_\shortparallel^2-m_f^2}(1-i\gamma_1\gamma_2)\right],
\eea
where `\textit{sfa}' indicates the strong field approximation and $\textsf{Tr}$ 
represents only the Dirac trace. Now one can notice that the longitudinal and 
transverse parts are completely separated and 
the gaussian integration over the transverse momenta can be done trivially, which leads to
\bea
\Pi_{\mu\nu}(p)\Big\vert_{sfa} &=& -iN_c\sum_{f}~e^{{-p_\perp^2}/{2q_fB}}~~\frac{q_f^3 B}{\pi}\int\frac{d^2k_\shortparallel}{(2\pi)^2} 
\frac{S_{\mu\nu}}{(k_\shortparallel^2-m_f^2)(q_\shortparallel^2-m_f^2)}, 
\label{pol_vacuum}
\eea
with the tensor structure $S_{\mu\nu}$ that originates from the Dirac trace  is 
\bea
S_{\mu\nu} = k_\mu^\shortparallel q_\nu^\shortparallel + q_\mu^\shortparallel k_\nu^\shortparallel 
- g_{\mu\nu}^\shortparallel \left((k\cdot q)_\shortparallel -m_f^2\right),
\eea
where the Lorentz indices  $\mu$ and $\nu$ are restricted to longitudinal values 
and forbids to take any transverse values. 
In vacuum, (\ref{pol_vacuum}) can be 
simplified using the Feynman parametrization technique \cite{calucci}, after which 
the structure of the photon polarization tensor can be written in compact form as
\bea
\Pi_{\mu\nu}(p) = \left(\frac{p^\shortparallel_\mu p^\shortparallel_\nu}{p_\shortparallel^2}-g^\shortparallel_{\mu\nu}\right)\Pi (p^2), \nn
\eea
which directly implies that due to the current conservation, the two point function is transverse. 
The scalar function $\Pi (p^2)$ is given by, 
\bea
\Pi (p^2) &=& N_c\sum_{f}\frac{q_f^3B}{8\pi^2 m_f^2}\, e^{-{p_\perp^2}/{2q_fB}}\left[4m_f^2+\frac{8m_f^4}{p_\shortparallel^2}
\left(1-\frac{4m_f^2}{p_\shortparallel^2}\right)^{-{1}/{2}}
\ln\frac{\left(1-\frac{4m_f^2}{p_\shortparallel^2}\right)^{{1}/{2}}+1}{\left(1-\frac{4m_f^2}{p_\shortparallel^2}\right)^{{1}/{2}}-1}\right].
\label{vacuum_check}
\eea
We note that the lowest threshold (LT)  for a photon to decay into fermion and antifermion
is provided by the energy conservation when photon momenta $p_\shortparallel^2(=\omega^2-p_3^2)=(m_f+m_f)^2= 4m_f^2$.  
Interestingly $\Pi(p^2)$ is singular in presence of magnetic field at this threshold.  This is 
because of the appearance of the pre-factor $\sqrt{1-4m_f^2 /p_\shortparallel^2}$ in the denominator
of the second term in (\ref{vacuum_check}) due to the dimensional
reduction from (3+1) to (1+1) in presence of the strong magnetic field.
This behavior is in contrast to that in absence of the magnetic field where the similar prefactor 
appears in the numerator~\cite{peskin}.  
Now, we explore $\Pi(p^2)$ physically in the following two domains around the LT, $p_\shortparallel^2=4m_f^2$ :

\begin{enumerate}

\item 
{\sf {Region-I}}~~$p_\shortparallel^2 < 4m_f^2$~:~~ In this case with 
$a=\sqrt{4m_f^2 /p_\shortparallel^2-1}$, 
let us write the logarithmic term in the second term of  (\ref{vacuum_check})  as
\bea
\ln\left(\frac{ai+1}{ai-1}\right) = \ln\left(\frac{re^{i\theta_1}}{re^{i\theta_2}}\right) = i(\theta_1-\theta_2),
\eea
where $r=\sqrt{(1+a^2)}$, $\theta_1 = \arctan(a)$ and $\theta_2 = \arctan(-a)$. 
Thus in (\ref{vacuum_check}) the logarithmic term is purely 
imaginary but overall $\Pi(p^2)$  is real because of the prefactor 
$\left(1-{4m_f^2}/{p_\shortparallel^2}\right)^{-{1}/{2}}$ being imaginary. 
Even if we choose the 
limit $p_\shortparallel^2<0$, then also the whole term is real again, since the denominator of the logarithmic term, 
$\sqrt{1-4m_f^2 /p_\shortparallel^2}$, is always greater than unity. So in 
the region $p_\shortparallel^2 < 4m_f^2$, $\Pi(p^2)$ is purely real. 
 
\item {\sf{Region-II}}~~$p_\shortparallel^2 > 4m_f^2$~:~~
Though in this limit the prefactor is real definite, but the denominator in the logarithmic term becomes negative and a complex number arises 
from it as $\ln(-x)=\ln\vert x\vert+i~\pi$. Thus we get both real and imaginary 
contributions, {\it i.e}, ${\mathcal{R}}e~\Pi(p^2)$ and $\mathcal{I}m~\Pi(p^2)$,
in this limit. The imaginary contribution is  relevant for studying the spectral function and its spectral properties.
\end{enumerate}

\begin{figure}
\begin{center}
\includegraphics[scale=0.85]{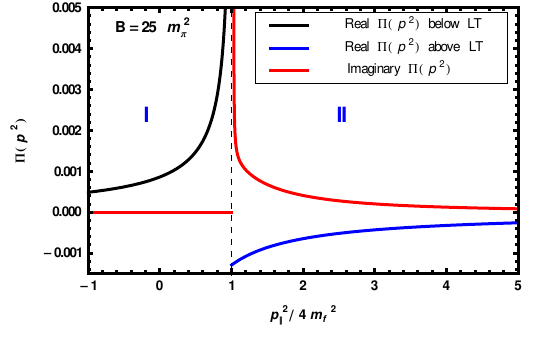}
\includegraphics[scale=0.85]{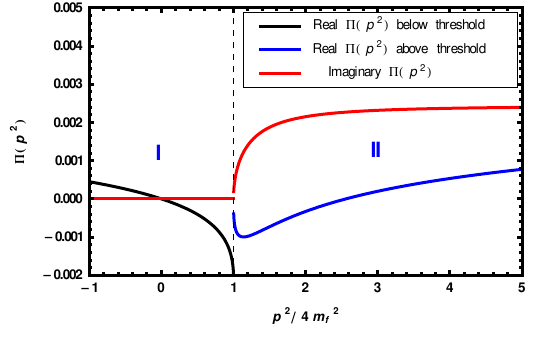}
\end{center}
\caption{ Plot of real and imaginary parts of $\Pi(p^2)$ as a function scaled photon momentum square with respect to
LT in various kinematic regions I and II as discussed in the text  in presence of a 
strong magnetic field (left panel) and in absence
of a magnetic field (right panel).}
\label{pola_plot}
\end{figure}

We now extract the vacuum spectral function in presence of strong magnetic field 
following (\ref{spec_func}) as
\bea
\rho\Big\vert_{sfa}^{\textsf{vacuum}} &=& \frac{1}{\pi} 
\mathcal{I}m~C^\mu_\mu(p)\Big\vert_{sfa}^{\textsf{vacuum}} = 
N_c\sum_{f}\frac{q_fBm_f^2}{\pi^2 p_\shortparallel^2}\ 
e^{-{p_\perp^2}/{2q_fB}}~\Theta\left(p_\shortparallel^2-4m_f^2\right)
\left(1-\frac{4m_f^2}{p_\shortparallel^2}\right)^{-{1}/{2}}. \label{spec_vac}
\eea
As seen the imaginary part is restricted by the LT, 
$p_\shortparallel^2=4m_f^2$. Below this threshold 
($p^{2}_\shortparallel<4m_f^2$), $\Pi(p^2)$ is real and there is no 
electromagnetic spectral contribution in vacuum with strong magnetic field as 
can be seen from region I in the left panel of Fig.~\ref{pola_plot}. This 
implies that there 
is also no creation of particle and antiparticle in vacuum below LT because the width
of the electromagnetic spectral function vanishes. 
Beyond LT there is also a continuous contribution (blue solid line in 
region II) in real part of $\Pi(p^2)$.
As seen the real part of $\Pi(p^2)$ is continuous both below and above the 
LT but has a discontnuity at the LT, $p^2_\shortparallel=4m_f^2$. 
Though we are interested in the imaginary part, we want to note that the real part can 
be 
associated with the dispersion property of vector boson~\footnote{This  has 
been discussed in Refs.~\cite{aminul,najmul} without  magnetic field and in 
Ref.~\cite{gusynin} with magnetic field.}. 
On the  other hand  the imaginary part
of the electromagnetic polarization tensor is associated with
interesting spectral properties of the system. So, beyond the LT 
($p^2_\shortparallel> 4m_f^2$) there is nonzero continuous contribution 
to the electromagnetic spectral function as given by (\ref{spec_vac}) and represented by a 
red solid line in region II in the left
panel of Fig.~\ref{pola_plot}. The right panel of Fig.~\ref{pola_plot} 
displays the analytic structure of vacuum $\Pi(p^2)$ in absence
of magnetic field~\cite{peskin}. In particular the comparison of the imaginary part of 
$\Pi(p^2)$ in absence of the magnetic field with that in
presence of the strong  magnetic field reveals an opposite trend  around LT. This is due 
to the effect of dimensional reduction in presence 
of the strong magnetic field. As a consequence the imaginary part of $\Pi(p^2)$ in 
presence of 
strong magnetic field would provide a very strong width to 
the photon that decays into particle and antiparticle, vis-a-vis an enhancement of the 
dilepton production from the hot and dense medium produced
in  heavy-ion collisions. So far we have discussed some aspects of the electromagnetic 
polarization tensor with a strong background magnetic field in vacuum.
Now we extend this to explore the spectral properties of a medium created in heavy-ion 
collisions with a strong background magnetic field. 

In the present situation without any loss of information we can contract the indices $\mu$ and $\nu$ in (\ref{pol_vacuum}), 
thus resulting in a further simplification as
\bea
\Pi_\mu^\mu(p)\Big\vert_{sfa} &=&  -iN_c\sum_{f}~e^{{-p_\perp^2}/{2q_fB}}~~\frac{q_f^3B}{\pi}\int\frac{d^2k_\shortparallel}{(2\pi)^2} 
\frac{2m_f^2}{(k_\shortparallel^2-m_f^2)(q_\shortparallel^2-m_f^2)}. 
\eea
At finite temperature this can be written by replacing the $p_0$ integral by Matsubara sum as 
\bea
\Pi_\mu^\mu(\omega,{\bf p})\Big\vert_{sfa} &=& -iN_c\sum_{f}~e^{{-p_\perp^2}/{2q_fB}}~~\frac{2q_f^3Bm_f^2}{\pi}
\left(i T \sum_{k_0}\right)\int\frac{dk_3}{2\pi} 
\frac{1}{(k_\shortparallel^2-m_f^2)(q_\shortparallel^2-m_f^2)} \, . 
\label{pola_ft}
\eea
We now perform the Matsubara sum using the mixed representation prescribed by 
Pisarski~\cite{pisarski}, where the trick is to dress 
the propagator in a way, such that it is spatial in momentum representation, but temporal in co-ordinate representation: 
\bea
\frac{1}{k_\shortparallel^2-m_f^2} \equiv \frac{1}{k_0^2-E_k^2} = \int\limits_0^\beta d\tau e^{k_0\tau} \Delta_M(\tau,k),
\label{mixed_representation}
\eea
 and 
\bea
\Delta_M(\tau,k) = \frac{1}{2E_k}\left[\left(1-n_F(E_k)\right)e^{-E_k\tau}-n_F(E_k)e^{E_k\tau}\right],
\eea
where $E_k=\sqrt{k_3^2+m_f^2}$ and $n_F(x)=(\exp(\beta x)+1)^{-1}$ is the Fermi-Dirac distribution function
with $\beta=1/T$. Using these, (\ref{pola_ft})
can be simplified as
\bea
\Pi_\mu^\mu(\omega,{\bf p})\Big\vert_{sfa} 
&=& N_c\sum_{f}e^{\frac{-p_\perp^2}{2q_fB}}~~\frac{2q_f^3Bm_f^2}{\pi}
T \sum_{k_0}\int\frac{dk_3}{2\pi} \int\limits_0^\beta~d\tau_1\int\limits_0^\beta~d\tau_2~ e^{k_0\tau_1}~e^{(k_0-p_0)\tau_2}\Delta_M(\tau_1,k)\Delta_M(\tau_2,q)\nn\\
&=& N_c\sum_{f}e^{\frac{-p_\perp^2}{2q_fB}}~~\frac{2q_f^3Bm_f^2}{\pi}\int\frac{dk_3}{2\pi}\int\limits_0^\beta~d\tau~ e^{p_0\tau}~\Delta_M(\tau,k)\Delta_M(\tau,q).
\eea
Now the $\tau$ integral is trivially performed as
\bea
\Pi_\mu^\mu(\omega,{\bf p})\Big\vert_{sfa}\!\! \!\! &=& \! \! 
N_c\sum_{f}e^{\frac{-p_\perp^2}{2q_fB}}~~\frac{2q_f^3Bm_f^2}{\pi}\int\frac{dk_3}{2\pi}
\sum_{l,r=\pm 1}\!\!
\frac{\left(1-n_F(rE_k)\right)\left(1-n_F(lE_q)\right)}{4(rl)E_kE_q(p_0-rE_k-lE_q)}\left[e^{-\beta(rE_k+lE_q)}-1\right] .
\label{Pi_sfa}
\eea
One can now easily  read off the discontinuity using
\bea
\textsf{Disc~}\left[\frac{1}{\omega +\sum_i E_i}\right]_\omega = - \pi\delta(\omega + \sum_i E_i),
\label{disc_delta}
\eea
which leads to
\bea
\mathcal{I}{ m}\, \Pi_\mu^\mu(\omega,{\bf p})\Big\vert_{sfa}\!\! \!\! &=& \! \! 
-N_c \pi 
\sum_{f}e^{\frac{-p_\perp^2}{2q_fB}}~~\frac{2q_f^3Bm_f^2}{\pi}\int\frac{dk_3}{2\pi}
\sum_{l,r=\pm 1}\!\!
\frac {\left(1-n_F(rE_k)\right)\left(1-n_F(lE_q)\right)}
{4(rl)E_kE_q} \nn \\
&& \times \left[ e^{-\beta(rE_k+lE_q)}-1\right]
\delta(\omega-rE_k-lE_q).
\label{Pi_sfa_gen}
\eea

The general form of the delta function in (\ref{Pi_sfa_gen})  corresponds to four 
processes\footnote{For LLL we have explicitly checked that these four 
processes can also be seen from (4.19) in Ref.~\cite{sadooghi} that uses Ritus method.}
for the choice of $r=\pm 1$ and $l=\pm 1$ as discussed below:
\begin{enumerate}
 \item  $r=-1$ and $l=-1$ corresponds to a process with  $\omega <0$, which 
violates energy conservation as all the quasiparticles 
have positive energies.
\item (a)  $r=+1$ and $l=-1$ corresponds to a process, $q\rightarrow q\gamma$, where 
a quark with energy $E_k$ makes a transition to an energy $E_q$ after emitting a  
timelike photon of energy $\omega$. (b) $r=-1$ and $l=1$ corresponds to similar case as 
(a). It has 
explicitly been shown in Appendix~\ref{app_a} that both processes are not allowed by the 
phase space and the energy conservation. In other words, the production of a timelike 
photon from one loop photon polarization tensor  is forbidden by the phase space and the 
energy conservation. 
However, we note here that these processes are somehow found to be 
nonzero for LLL in Ref.~\cite{sadooghi}.
\item $r=1$ and $s=1$ corresponds to a process where a quark and a antiquark 
annihilate to a virtual photon, which is the only allowed process:
\end{enumerate}
So, for the last case, one can write from (\ref{Pi_sfa_gen}) 
\bea
\mathcal{I}m~\Pi_\mu^\mu(\omega,{\bf p})\Big\vert_{sfa} &=& 
N_c \pi \sum_{f}
e^{\frac{-p_\perp^2}{2q_fB}}~~\frac{2q_f^3Bm_f^2}{\pi}
\int\frac{dk_3}{2\pi}~\delta(\omega-E_k-E_q)\frac{\left[1-n_F(E_k)-n_F(E_q)\right]}{
4E_kE_q}.
\label{Pi_im}
\eea

After performing the $k_3$ integral using (\ref{a3}) the spectral function  in strong 
field approximation  is finally obtained following (\ref{spec_func}) as
\bea
\rho \Big\vert_{sfa}&=&\frac{1}{\pi}\mathcal{I}m~ 
C^\mu_\mu (p)\Big\vert_{sfa}\nn\\
&=& N_c\sum_{f}\frac{q_fBm_f^2}{\pi^2 p_\shortparallel^2}~e^{-{p_\perp^2}/{2q_fB}}~\Theta
\left(p_\shortparallel^2-4m_f^2\right)\left(1-\frac{4m_f^2}{p_\shortparallel^2}
\right)^{-{1}/{2}} \Bigl[1-n_F(p_+^s)-n_F(p_-^s)\Bigr],
\label{spec_sfa_general}
\eea
where 
\bea
p_\pm^s = \frac{\omega}{2}\pm \frac{p_3}{2}\sqrt{\left(1-\frac{4m_f^2}{p_\shortparallel^2}\right)}.
\eea

We note that the electromagnetic spectral function in strong field approximation obtained 
here in (\ref{spec_sfa_general}) using Schwinger method has a factor 
$[1-n_F(p_+^s)-n_F(p_-^s)]$. This thermal factor appears when a quark 
and antiquark annihilate to a virtual photon in a thermal medium, which is the only 
process  allowed by the phase space as shown in our calculation. In Ref.\cite{sadooghi}
besides this, there also appears additional thermal factors due to the presence 
of the transition processes ($q\rightarrow q\gamma$) as discussed above in 2(a) and 
2(b) and in Appendix~\ref{app_a}.

The vacuum part in presence of the strong magnetic field can be easily separated out from (\ref{spec_sfa_general}) as
\bea
\rho \Big\vert_{sfa}^{\textsf{vacuum}} &=& N_c\sum_{f}\frac{q_fBm_f^2}{\pi^2 
p_\shortparallel^2}~e^{-{p_\perp^2}/{2q_fB}}~\Theta
\left(p_\shortparallel^2-4m_f^2\right)\left(1-\frac{4m_f^2}{p_\shortparallel^2}\right)^{-{1}/{2}},
\eea
which agrees with that obtained in (\ref{spec_vac}).

We outline some of the important features of the spectral functions:

\begin{enumerate}
\item[(i)] In general the electromagnetic spectral function in (\ref{spec_sfa_general})  vanishes in the massless limit of  quarks. 
This particular feature arises because of the presence of magnetic field which  reduces the system  to ($1+1$) dimension. This can 
be further understood from the symmetry argument and is attributed to the CPT invariance of the theory~\cite{Das:2012pd}. Physically 
this observation further signifies that in ($1+1$) dimension an on-shell massless thermal fermion cannot scatter in the forward direction.
 
\item[(ii)] The threshold, $p_\shortparallel^2=4m_f^2$,  for LLL is  independent of the magnetic field 
strength. It is also independent of $T$ as $q_fB \gg T^2$ in the strong field approximation. 
Like vacuum case here also
the spectral function vanishes below the threshold and there is no pair creation of 
particle and antiparticle. 
This is because the polarization tensor is purely real below the threshold. This implies that the momentum of the external photon supplies energy and 
virtual pair in LLL becomes real via photon decay. 

\item[(iii)] When the photon longitudinal momentum square is equal to the LT,  $p_\shortparallel^2=4m_f^2 $, it strikes the LLL and 
the spectral strength diverges because of the factor 
$\left(1-{4m_f^2}/{p_\shortparallel^2}\right)^{-{1}/{2}}$ that appears due to 
the dimensional reduction. 
Since the LLL dynamics is (1+1) dimensional, 
there is a dynamical mass generation~\cite{gusynin, ferrer} of the 
fermions through mass operator (\textit{e.g.} chiral condensate), 
which causes the magnetic field induced chiral symmetry breaking
in the system. This suggests that the strong fermion pairing takes place
in LLL~\cite{gusynin} even at the weakest attaractive interaction between fermions in 
(3+1) dimension. A (3+1) dimensional weakly interacting system  
in presence of strong magnetic 
field can be considered as a strongly correlated system 
in LLL dynamics which is (1+1) dimensional.
In that case 
$m_f$ should be related to the dynamical mass provided by the 
condensates~\cite{gusynin,ferrer}. One can incorporate it based on nonperturbative
model calculations, then LT will change accordingly.

\item[(iv)] The spectral strength starts with a high value for the photon longitudinal 
momentum $p_\shortparallel > 2m_f$  due to the dimensional reduction  
or LLL dynamics and then falls off with increase of $\omega$ as there is nothing beyond 
the LLL in strong field approximation. To improve the high energy behavior of the 
spectral function  one requires weak field approximation ($T^2~\gg~q_fB$).
\end{enumerate}

\begin{figure}
\begin{center}
\includegraphics[scale=0.8]{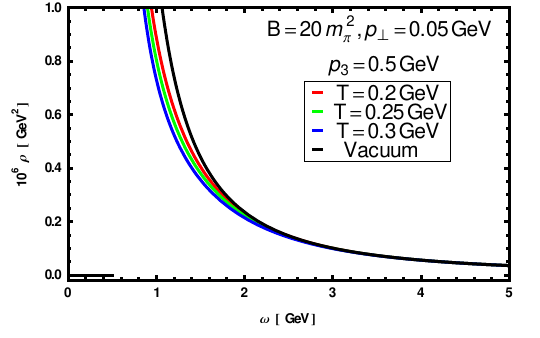}
\hspace{1cm}\includegraphics[scale=0.8]{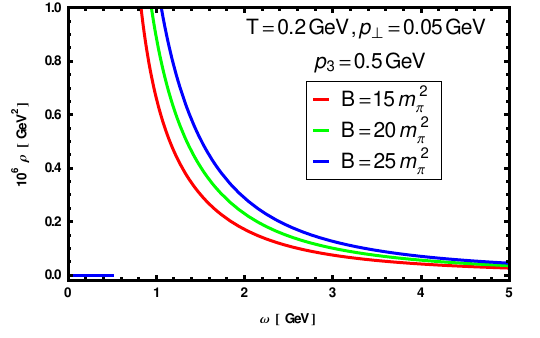}
\end{center}
\caption{{\it Left panel:} Variation of the spectral function with photon energy for different values of $T$ at fixed $B$, $p_\perp$ and $p_3$.
{\it Right panel:} Same as left panel but for different values of magnetic field  at fixed $T$, $p_\perp$ and $p_3$.The value of the magnetic 
field is chosen in terms of the pion mass $m_\pi$.}
\label{spec_plot_1}
\end{figure}

In Fig. \ref{spec_plot_1} the variation of the spectral function with photon energy $\omega$ for different values of $T$ in the left panel and for
different values of magnetic field in the right panel. With increase in $T$  the spectral strength in the left panel gets depleted because of the presence 
of the thermal weight factor $[1-n_F(p_+^s)-n_F(p_-^s)]$ as the distribution functions $n_F(p_\pm^s)$ increase with $T$ that restricts the available 
phase space. Nevertheless the effect of temperature is small in the strong field 
approximation as $q_fB\gg T^2$.  On the other hand the spectral
strength in the right panel increases with the increase of the magnetic field $B$ as the 
spectral function is proportional to $B$. 

\begin{figure}
\begin{center}
\includegraphics[scale=0.8]{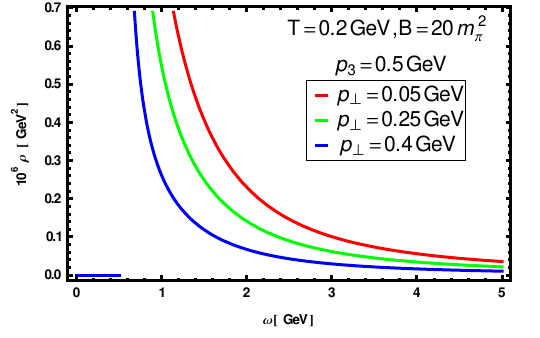}
\end{center}
\caption{Variation of the spectral function with photon energy $\omega$ for different values of transverse momentum at fixed $B$, $T$ and  $p_3$.}
\label{spec_plot_2}
\end{figure}

In Fig.~\ref{spec_plot_2} the variation of the spectral function with photon energy 
$\omega$  is shown for three different values of 
the transverse momentum $p_\perp$. The spectral function is found to get 
exponentially suppressed with 
the gradually increasing value of $p_\perp$.

We also consider a special case where the external three momentum ($p$) of photon is 
taken to be zero and the simplified expression for the spectral function comes out to be,
\bea
\rho(\omega)\Big\vert_{sfa}=\frac{1}{\pi}\mathcal{I}m~ C^\mu_\mu (\omega,{\bf 
p}=0)\Big\vert_{sfa} &=& N_c\sum_{f}\frac{q_fBm_f^2}{\pi^2 \omega^2}
\Theta\left(\omega^2-4m_f^2\right)
\left(1-\frac{4m_f^2}{\omega^2}\right)^{-{1}/{2}} \Bigl[1-2 n_F\left(\frac{\omega}{2}\right)\Bigr].
\label{spec_sfa}
\eea

\begin{figure}
\begin{center}
\includegraphics[scale=0.8]{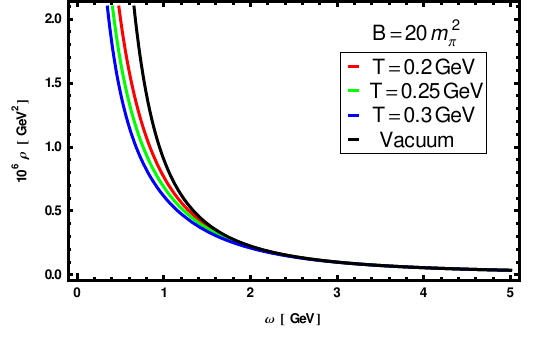}
\hspace{1cm}\includegraphics[scale=0.8]{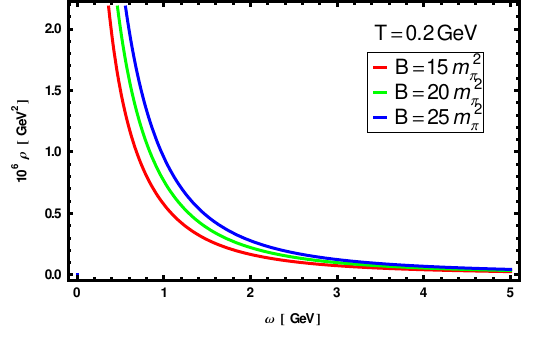}
\end{center}
\caption{Same as Fig.~\ref{spec_plot_1} but for zero external three momentum ($p$) of photon.}
\label{spec_plot_3}
\end{figure}

In Fig.~\ref{spec_plot_3} same things are plotted as in Fig.~\ref{spec_plot_1} but for a 
simplified case of zero external three momentum of photon. As can be seen from 
(\ref{spec_sfa}), here the value of the threshold is shifted to photon energy as 
$\omega=2m_f$ and the shape of the plots are slightly modified. In the following 
subsec.~\ref{dil} as a spectral property we discuss the leading order thermal dilepton 
rate for a magnetized medium.

\section{Dilepton rate}
\label{dil}
\subsection{ Dilepton rate in absence of external magnetic field}
\label{dil_unmag}
The dilepton multiplicity per unit space-time volume is given~\cite{Weldon:1990iw} as
\bea
\frac{dN}{d^4x}&=& 2\pi e^2 e^{-\beta 
p_0}L_{\mu\nu}\rho^{\mu\nu}\frac{d^3{\mathbf 
q}_1}{(2\pi)^3E_1}\frac{d^3{\mathbf q}_2}{(2\pi)^3E_2}, \label{d1}
\eea
where ${\mathbf q}_i$ and $E_i$ with $i=1,2$ are three momentum and energy 
lepton pairs.
The photonic tensor or the electromagnetic spectral function can be written as
\bea
\rho^{\mu\nu}(p_0,{\bf p}) &=& -\frac{1}{\pi}\frac{e^{\beta p_0}}{e^{\beta 
p_0}-1}{\cal I}m\left[D_R^{\mu\nu}(p_0,{\bf p})\right]\equiv 
-\frac{1}{\pi}\frac{e^{\beta p_0}}{e^{\beta 
p_0}-1}~\frac{e_e^2}{p^4}{\cal I}m\left[C^{\mu\nu}(p_0,{\bf p})\right], \label{d2}
\eea
where $e_e$ is the relevant electric charge, $C^{\mu\nu}$ is the two 
point current-current correlation function, whereas 
$D_R^{\mu\nu}$ represents the photon propagator. Here we used the 
relation~\cite{Weldon:1990iw} 
 \bea
e_e^2C^{\mu\nu} = p^4 D_R^{\mu\nu},
\label{d2i}
 \eea
where $e_e$ is the effective coupling. 

Also  the leptonic tensor in terms of Dirac spinors is given by
\bea
L_{\mu\nu} &=& \frac{1}{4} 
\sum\limits_{\mathrm{spins}}\mathrm{tr}\left[\bar{u}(q_2)\gamma_\mu 
v(q_1)\bar{v}(q_1)\gamma_\nu u(q_2)\right]
= q_{1\mu}q_{2\nu}+q_{1\nu}q_{2\mu}-(q_1\cdot q_2+m_l^2)g_{\mu\nu},
\label{d3}
\eea
where $q_i\equiv (q_0, {\mathbf q}_i)$ is the four momentum of the $i$th lepton.
Now inserting $\int d^4p\, \delta^4(q_1+q_2-p)=1$, one can write the dilepton 
multiplicity as
\bea
\frac{dN}{d^4x}\
&=& 2\pi e^2 e^{-\beta p_0}\int d^4p \, \delta^4(q_1+q_2-p)
L_{\mu\nu}\rho^{\mu\nu}\frac{d^3q_1}{(2\pi)^3E_1}\frac{d^3q_2}{ 
(2\pi)^3E_2}. \label{d4} 
\eea
Using the identity
\bea
\int\frac{d^3q_1}{E_1}\frac{d^3q_2}{E_2}
\delta^4(q_1+q_2-p)L_{\mu\nu} &=& 
\frac{2\pi}{3} 
\left(1+\frac{2m_l^2}{p^2}\right)\left(1-\frac{4m_l^2}{p^2}\right)^{1/2}
\left(p_\mu p_\nu-p^2g_{\mu\nu}\right) \nn\\
&=&\frac{2\pi}{3}F_1(m_l,p^2)\left(p_\mu p_\nu-p^2g_{\mu\nu}\right), \label{d5}
\eea
the dilepton production rate comes out to be,
\bea
\frac{dN}{d^4xd^4p} &=&
\frac{\alpha_{\mbox{em}}e_{e}^2} 
{12\pi^3}\frac{n_B(p_0)}{p^2} F_1(m_l,p^2) \left(\frac{1}{\pi} {\cal I}m \left 
[C^{\mu}_{\mu}(p_0,{\bf p})\right]\right), \label{d6}
\eea 
where $n_B(p_0)=(e^{p_0/T}-1)^{-1}$.
Now if we consider a two-flavor case, $N_f=2$,
\bea
e_e^2 &=&  \sum\limits_f q_f^2 = \frac{5}{9}e^2 = 
\frac{5\times 
4\pi\alpha_{\mbox{em}}}{9}, \label{d7}
\eea
and the dilepton rate can be written as
\bea
\frac{dN}{d^4xd^4p} &=& \frac{5\alpha_{\mbox{em}}^2}{27\pi^2} 
\frac{n_B(p_0)}{p^2} F_1(m_l,p^2) 
\left(\frac{1}{\pi}\mathrm{Im} \left 
[C^{\mu}_{\mu}(p_0,{\bf p})\right]\right),
\label{d8}
\eea
where the invariant mass of the lepton pair $M^2\equiv p^2(=p_0^2-|{\mathbf 
p}|^2=\omega^2-|{\mathbf p}|^2)$. We 
note that for massless lepton ($m_l=0$) $F_1(m_l,p^2)=1$. 

The above expression applies to a system in the absence of magnetic fields. In its presence, one cannot factor out $q_f^2$, because of difference in coupling strengths of $u$ and $d$ quarks to the magnetic fields. Thus, in the presence of magnetic fields, Eq.~(\ref{d6}) is to be applied carefully as shown below.

\subsection{Dilepton rate in presence of strong external constant magnetic field}
\label{dil_mag}

We first would like to note that the dileptons are produced in all stages of 
the hot and dense fireball created in heavy-ion collisions. They are produced in 
leading order from the decay of a virtual photon through the annihilation of 
quark-antiquark pairs. In  
non-central heavy-ion collisions an anisotropic magnetic field is 
expected to be generated in the direction perpendicular to the reaction plane, 
due to the relative motion of the heavy-ions themselves (spectators). It is believed that 
the
initial magnitude of this magnetic field can be very high at
the time of the collision and then it decreases very 
fast~\cite{bzdak,McLerran}. 
The dilepton production from a magnetized hot and dense matter can generally be dealt 
with three different scenarios~\cite{Tuchin:2013prc,sadooghi}: 
(1) only the quarks move in a magnetized medium but not the final lepton pairs, 
(2) both quarks and leptons move in a magnetized medium and  (3) only the final lepton 
pairs move in the magnetic field. 

\subsubsection{Quarks move in a strong magnetized medium but not the final lepton pairs}

We  emphasize that the case  we consider here is interesting and very much relevant to 
noncentral heavy-ion collisions, especially for the scenario of fast decaying magnetic 
field~\cite{bzdak,McLerran} and also for lepton pairs produced late or at the edges 
of hot and dense magnetized medium so that they are unaffected by the magnetic field. 
In this scenario only the electromagnetic spectral function $\rho^{\mu\nu}$ in (\ref{d1}) 
 will be modified by the background constant magnetic field whereas the leptonic tensor 
$L_{\mu\nu}$ and the phase space factors will remain unaffected. The dilepton rate for 
massless ($m_l=0$) leptons can then be written from (\ref{d6}) as
\begin{eqnarray}
    \frac{dN}{d^4xd^4p} &=& \frac{\alpha_{\rm em}}{12 \pi^3}\frac{n_B(p_0)}{p^2}\sum_f q_f^2 \left [\rho_f (p_\shortparallel,p_\perp)\right]_m \nonumber\\   
    &=&\frac{N_c \alpha_{\rm em}}{12\pi^5} n_B(\omega)\sum_f \frac{q_f^2|q_fB| m_f^2}{p^2p_\shortparallel^2}~e^{-{p_\perp^2}/{2|q_fB|}}~\Theta\left(p_\shortparallel^2-4m_f^2\right)\left(1-\frac{4m_f^2}{p_\shortparallel^2}\right)^{-{1}/{2}}\\ \nonumber
    &&\times\Bigl[1-n_F(p_+^s)-n_F(p_-^s)\Bigr], \label{d9}
\end{eqnarray}
where the electromagnetic spectral function  $[\rho_f (p_\shortparallel,p_\perp)]_m$ 
in hot
magnetized medium has been used from (\ref{spec_sfa_general}) and the subscript $f$ is used to denote its flavor dependence. The invariant mass 
of the lepton pair is $M^2 \equiv p^2 (\omega^2-\vert \mathbf{p}\vert^2) = 
\omega^2-p_3^2-p_\perp^2=p_\shortparallel^2-p_\perp^2$.

\begin{figure}
\begin{center}
\includegraphics[scale=0.4]{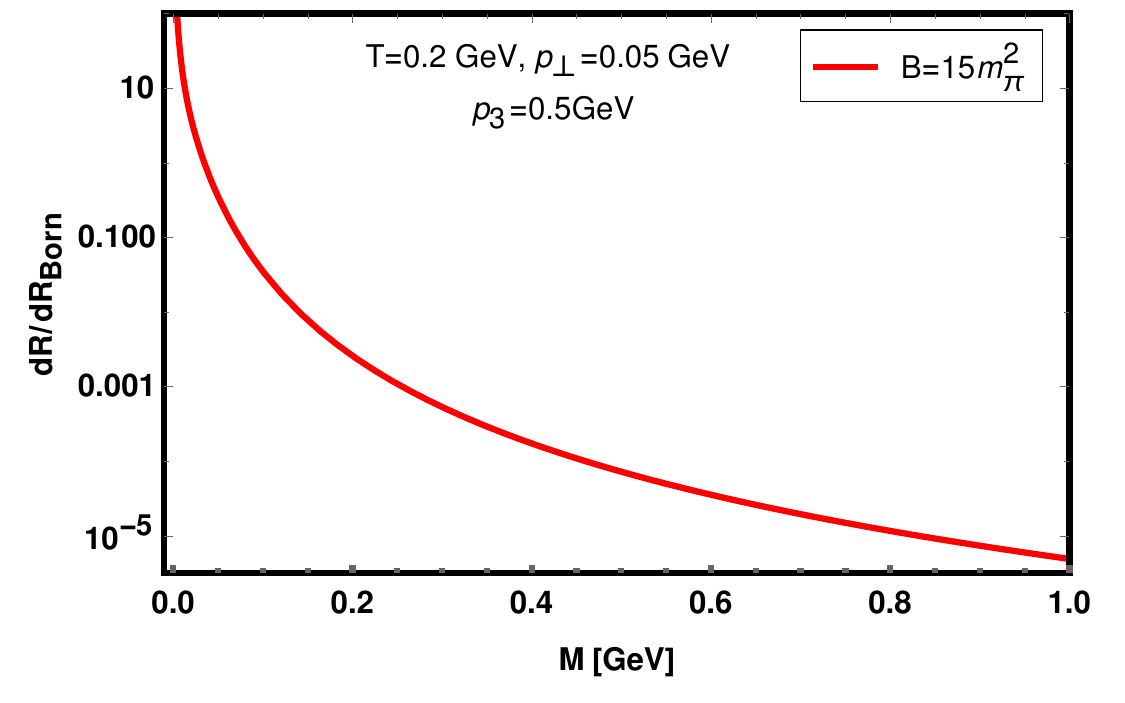}
\hspace{1cm}\includegraphics[scale=0.4]{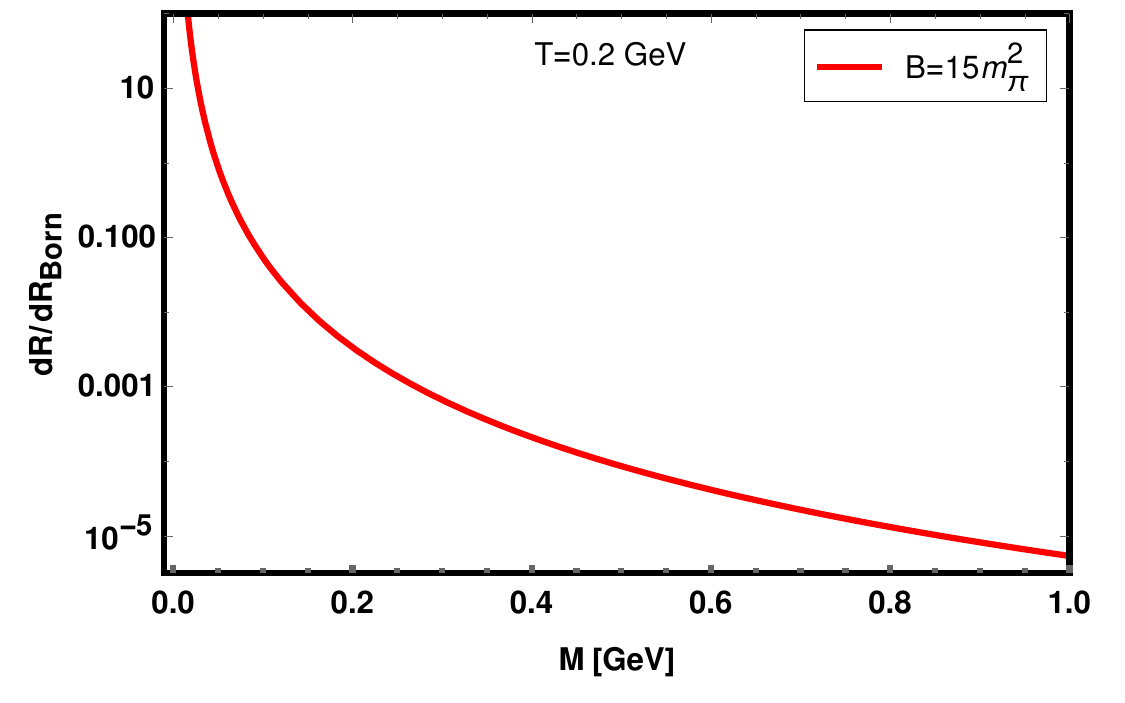}
\end{center}
\caption{Plot of ratio of the Dilepton rate in the strong magnetic field 
approximation to the Born rate (perturbative leading order) for both finite (left panel)  
and zero (right panel) external three momentum of photon.}
\label{dr_comparison}
\end{figure}

In Fig.~\ref{dr_comparison} a ratio of the dilepton rate in the present scenario with 
strong field approximation to that of the perturbative leading order (Born) dilepton rate 
is displayed as a function of the invariant mass. The left panel is for finite external 
photon momentum  
whereas the right panel is for zero external photon momenta. 
The features of the spectral function as 
discussed above are reflected in these dilepton rates. The LLL dynamics in 
strong field approximation enhances the dilepton rate as compared to 
the Born rate for a very low invariant 
mass ($\le 100 $ MeV), whereas at high mass it falls off very 
fast similar to that of the spectral function since there is no higher LL in
strong field approximation as noted in 
point (iv). One requires weak field approximation ($q_fB~<<~T^2$) to improve the
high mass behavior of the dilepton rate.
We note that the enhancement found in the strong field approximation in the 
rate will contribute to the dilepton 
spectra at low invariant mass, which is however beyond the scope of the present detectors 
involved in heavy-ion collisions 
experiments. 

\subsubsection{Both quark and lepton move in magnetized medium in strong field 
approximation}
This scenario  is expected to be the most general one. To consider such a 
scenario the usual dilepton production 
rate given in (\ref{d6}) has to be supplemented with the appropriate 
modification of the electromagnetic and leptonic tensor along with the phase space 
 factors in a magnetized medium.
Since we are interested in only LLL, 
we briefly outline below the required modification~\footnote{A  detailed calculation for 
more  general case is under progress.} in the dilepton production rate only 
for LLL:
\begin{enumerate}
\item[$\bullet$]
The phase space factor in presence of magnitized medium gets modified~\cite{landau}
as
\bea
\frac{d^3{\mathbf q}}{(2\pi)^3E} & \rightarrow& \frac 
{|eB|}{(2\pi)^2}\sum\limits _{n=0}^{\infty} \frac{dq_z}{E}. \label{d10}
\eea
where $d^2q_\perp=2\pi|eB|$, $e$ is the electric charge of the lepton and $\sum\limits 
_{n=0}^{\infty}$ is over 
LL. For for strong magnetic field one is 
confined in LLL and $n=0$ only. The factor $|eB|/(2\pi)^2$ is the density of states 
in 
the transverse direction and true for LLL~\cite{gusynin}.

\item[$\bullet$] The electromagnetic spectral function gets modified for LLL
as already been discussed in Sec.~\ref{sfa}.

\item[$\bullet$] In presence of constant magnetic field  the spin of 
fermions is aligned along the field direction and the usual 
Dirac spinors $u(q)$ and $v(q)$ in (\ref{d3}) get modified~\cite{schwinger,gusynin} 
by $P_n u(\tilde{q})$ and  $P_n v(\tilde{q})$  with  
$\tilde{q}^\mu=\left(q^0,0,0,q^3\right)$ and $P_n$ is the projection operator at the 
$n$th LL. For LLL  it takes a simple form
\bea
P_0 = \frac{1-i\gamma_1\gamma_2}{2}. \label{d11}
\eea
Now, the modification in the leptonic part in presence of a strong magnetic field 
can be carried out as
\bea
L^m_{\mu\nu} &=& \frac{1}{4} 
\sum\limits_{\mathrm{spins}}\mathrm{tr}\left[\bar{u}(\tilde{q}_2)P_0\gamma_\mu 
P_0v(\tilde{q}_1)\bar{v}(\tilde{q}_1)P_0\gamma_\nu P_0 u(\tilde{q}_2)\right]\nn\\
&=& \frac{1}{4} 
\mathrm{tr}\left[(\slashed{\tilde{q}}_1+m_l)\left(\frac{1-i\gamma_1\gamma_2}{2}
\right)\gamma_\mu 
\left(\frac{1-i\gamma_1\gamma_2}{2}\right)(\slashed{\tilde{q}}_2-m_l)\left(\frac{
1-i\gamma_1\gamma_2}{2}\right) 
\gamma_\nu\left(\frac{1-i\gamma_1\gamma_2}{2}\right)\right]\nn\\
&=& \frac{1}{2} 
\left[q^\shortparallel_{1\mu}q^\shortparallel_{2\nu}+q^\shortparallel_{1\nu}
q^\shortparallel_{2\mu}-((q_1\cdot q_2)_\shortparallel + 
m_l^2)(g_{\mu\nu}^\shortparallel-g_{\mu\nu}^\perp-g_{1\mu}g_{1\nu}-g_{2\mu}g_{2\nu})\right
].
\label{d12}
\eea
\item[$\bullet$] Requires an insertion $\int d^2p^\shortparallel \, 
\delta^2(q_1^\shortparallel+q_2^\shortparallel-p^\shortparallel)=1$.
\item[$\bullet$] Replacing\footnote{The authors of Ref.\cite{sadooghi} 
replaced 
$d^2p^\perp= V^{2/3}(\frac{eB}{2\pi})^2$, where $V$ is the volume. This led to a different 
normalization factor in the dilepton rate in Ref.~\cite{sadooghi}.}  $d^2p^\perp= 
2\pi|eB|$  and $d^4p= d^2p^\perp 
d^2p^\shortparallel$.

\item[$\bullet$] Making use of an identity:
\bea
2\pi|eB| \int\frac{dq_1^z}{E1}\int\frac{dq_2^z}{E2}
~\delta^2(q_1^\shortparallel+q_2^\shortparallel-p^\shortparallel) \ L^m_{\mu\nu}
&=& 4\pi
\frac{|eB|m_l^2}{(p_\shortparallel^2)^2}\left(1-\frac{4m_l^2}{p_\shortparallel^2}
\right)^ { - {1}/{2}} \left(p^\shortparallel_\mu 
p^\shortparallel_\nu-p_\shortparallel^2g^\shortparallel_{\mu\nu}\right) \nn\\  
&=&\frac{4\pi}{(p_\shortparallel^2)^2} 
F_2(m_l,p_\shortparallel^2)\left(p^\shortparallel_\mu 
p^\shortparallel_\nu-p_\shortparallel^2g^\shortparallel_{\mu\nu}\right) . \label{d13}
\label{d13}
\eea
\end{enumerate}

Putting all these together, we finally obtain the
dilepton production rate from (\ref{d1}) for LLL as
\bea
\frac{dN^m}{d^4xd^4p}
&=&\frac{\alpha_{em}e_e^2}{2\pi^3}\frac{n_B(p_0)}{p_\shortparallel^2p^4}F_2(m_l,
p_\shortparallel^2)\left(\frac{1}{\pi}{\cal I} m \left[C^{\mu}_{\mu}(p_\shortparallel,
p_\perp)\right]\right)_m, \label{d14}
\eea
and for  two-flavor case  ($N_f=2$) it becomes
\begin{eqnarray}
    \frac{dN^m}{d^4xd^4p} &=& \frac{\alpha_{\rm em}}{2\pi^3}\frac{n_B(p_0)}{p_\shortparallel^2p^4}|eB|m_l^2\left(1-\frac{4m_l^2}{p_\shortparallel^2}\right)^{-{1}/{2}}\sum_f q_f^2\left [\rho_f (p_\shortparallel,p_\perp)\right]_m \nonumber\\ 
    &=&\frac{N_c\alpha_{\rm em}}{2\pi^5} n_B(\omega) \sum_f \frac{ q_f^2\,|eB|\,|q_fB| m_f^2 m_l^2}{p_\shortparallel^4p^4} \Theta\left(p_\shortparallel^2-4m_l^2\right)\left(1-\frac{4m_l^2}{p_\shortparallel^2}\right)^{-{1}/{2}} \Theta\left(p_\shortparallel^2-4m_f^2\right)\left(1-\frac{4m_f^2}{p_\shortparallel^2}\right)^{-{ 1}/{2}} \nn \\
    &&\times e^{-{p_\perp^2}/{2|q_fB|}} \Bigl[1-n_F(p_+^s)-n_F(p_-^s)\Bigr],\label{d15}
\end{eqnarray}

We now note that the dilepton production rate in (\ref{d15}) is of ${\cal O}[|eB|^2]$ in presence of magnetic field $B$ due to the effective dimensional reduction~\footnote{ 
One factor of $|eB|$ comes from the leptonic part, whereas another factor of $|eB|$ comes out from the electromagnetic spectral function involving quarks.}.  This 
dimensional reduction also renders a factor $1/\sqrt{1-{4m_l^2}/{p_\shortparallel^2}}$ in the leptonic part $L^m_{\mu\nu}$ 
that provides another threshold $p_\shortparallel^2 \ge 4m_l^2$ 
in addition to that coming from electromagnetic part $ p_\shortparallel^2 \ge 
4m_f^2$.  In general the mass of fermions in a magnetized hot medium will be 
affected by both temperature and magnetic field. The thermal 
effects~\cite{kapusta,lebellack} can be considered
through thermal QCD and QED, respectively, for quark ($\sim g^2T^2$; $g$ is the QCD 
coupling)  and lepton ($\sim e^2T^2$) whereas the magnetic effect comes through the 
quantized LL ($2n|q_fB|$). However, in LLL ($n=0$), the magnetic effect to the mass 
correction vanishes in strong field approximation. Also in strong field approximation 
($|q_fB| \gg T$), there could be dynamical mass generation 
through chiral condensates~\cite{gusynin} of quark and antiquark leading to
magnetic field induced chiral symmetry breaking, which could 
play a 
dominant role. Nevetheless, the threshold will, finally, be determined by the effective 
mass $\tilde m =$ max(${ m}_l,{ m}_f$) as 
$\Theta\left(p_\shortparallel^2-4{\tilde m}^2\right)$ and the dilepton rate in LLL reads 
as
\begin{eqnarray}
    \frac{dN^m}{d^4xd^4p} &=&\frac{N_c\alpha_{\rm em}}{2\pi^5}\sum_f\frac{q_f^2\,|eB|\,|q_fB| m_f^2 m_l^2}{p_\shortparallel^4p^4}\Theta\left(p_\shortparallel^2-4{\tilde m}^2\right)\left(1-\frac{4m_l^2}{p_\shortparallel^2}\right)^{-{1}/{2}}\left(1-\frac{4m_f^2}{p_\shortparallel^2}\right)^{-{ 1}/{2}} \nn \\
    && \times e^{-{p_\perp^2}/{2|q_fB|}}\, n_B(\omega) \Bigl[1-n_F(p_+^s)-n_F(p_-^s)\Bigr],\label{d16}
\end{eqnarray}
where the kinametical factors agree but the prefactor 
$(10/\pi^4)$ and the thermal factor $n_B(\omega)[1-n_F(p_+^s)-n_F(p_-^s)\Bigr]$  
differ from those of Ref.~\cite{sadooghi} and the reasons for which are discussed in 
details earlier. This restricts one to make a quantitative comparison of the dilepton 
rate with that obtained in Ref.~\cite{sadooghi}. We further note that a comparison with 
the experimental results  or the results (dilepton spectra)  obtained by 
Tuchin~\cite{Tuchin:2013prc}
needs a space-time evolution of the dilepton rate 
in a hot magnetized medium produced in heavy-ion collision. A proper space-time 
evolution requires  hydrodynamic prescription in presence of magnetic field, 
which is indeed a difficult task and beyond the scope of this article.

We also note that  the production rate for case - (3) requires 
modification of the leptonic 
tensor in a magnetized medium but the electromagnetic one remains unmagnetized. Since 
this is a rare possibility, we  skip the discussion here but can easily be obtained.


\section{Debye Screening in a strong magnetic field approximation}
\label{ds}
In this section we further explore the Debye screening mass in strongly magnetized hot 
medium. In the static limit the Debye screening mass is obtained as
\bea
m_D^2=\Pi_{00}(\omega=0,\vert\vec{p}\vert\rightarrow 0).
\eea
Using (\ref{pol_vacuum}) we get 
\bea
\Pi_{00}\Big\vert^{sfa}_{\vert\vec{p}\vert=0,\omega\rightarrow 0} &=& 
N_c\sum_f\frac{q_f^3B}{\pi}\int\limits_0^{\infty}\frac{dk_3}{2\pi}~T\sum_{k_0}\frac{S_{00}
}{(k_\shortparallel^2-m_f^2)^2}\nn\\
&=& N_c\sum_f\frac{q_f^3B}{\pi}\int\limits_0^{\infty}\frac{dk_3}{2\pi}~\left 
[\frac{1}{4\pi i}\oint dk_0\frac{S_{00}\left[1-2n_F(k_0)\right]}{(k_0^2-E_k^2)^2}\right ],
\label{pi00sfa}
\eea
where, $E_k^2 = k_3^2+m_f^2$ and at the limit of zero external three momentum and 
vanishing external energy $S_{00}$ comes out to be
\bea
S_{00}&=&k_0q_0+k_3q_3+m_f^2\Big\vert_{\vert\vec{p}\vert=0,\omega\rightarrow 0}\nn\\
&=&k_0^2+k_3^2+m_f^2,\nn\\
&=& (k_0^2-E_k^2)+2E_k^2. 
\eea
Now, the $k_0$ integration can be divided into two parts as
\bea
I_1 &=& \frac{1}{4\pi i}\oint dk_0\frac{\left[1-2n_F(k_0)\right]}{(k_0^2-E_k^2)}\nn\\
&=& \frac{1-2n_F(E_k)}{2E_k}, \\ 
\textsf{and~}I_2 &=& \frac{1}{4\pi i}\oint 
dk_0\frac{2E_k^2\left[1-2n_F(k_0)\right]}{(k_0^2-E_k^2)^2}\nn\\
&=& 
2E_k^2 \frac{d}{dk_0}\left(\frac{1-2n_F(k_0)}{(k_0+E_k)^2}\right)\Big\vert_{k_0=E_k}\nn\\
&=& -\frac{1-2n_F(E_k)}{2E_k}+\beta n_F(E_k)\left[1-n_F(E_k)\right].\\
\therefore I_1+I_2 &=& \beta n_F(E_k)\left[1-n_F(E_k)\right].
\eea
From (\ref{pi00sfa}) the temporal part of the polarization tensor in the limit of zero 
external three momentum (the long wavelength limit) and vanishing external energy comes 
out to be
\bea
\Pi_{00}\Big\vert^{sfa}_{\vert\vec{p}\vert=0,\omega\rightarrow 0} &=& 
N_c\sum_f\frac{q_f^3B}{\pi T}\int\limits_0^{\infty}\frac{dk_3}{2\pi}~ 
n_F(E_k)\left[1-n_F(E_k)\right].
\label{dsmassive}
\eea
For massive case ($m_f \neq 0$) this expression cannot be reduced further, analytically, 
by performing the $k_3$ integration. We evaluate it numerically to extract the essence of 
Debye screening. On the other hand, for the massless case ($m_f = 0$) a simple analytical 
expression is obtained as
\bea
\Pi_{00}\Big\vert^{sfa}_{\vert\vec{p}\vert,m_f=0,\omega\rightarrow 0} &=& 
N_c\sum_f\frac{q_f^3B}{\pi T}\int\limits_0^{\infty}\frac{dk_3}{2\pi}~ 
n_F(k_3)\left[1-n_F(k_3)\right],\nn\\
&=& N_c\sum_f\frac{q_f^3B}{\pi T} ~\frac{T}{4\pi}
= N_c\sum_f\frac{q_f^3B}{4\pi^2}.
\label{dsmassless}
\eea 

\begin{figure}
\begin{center}
\includegraphics[scale=0.8]{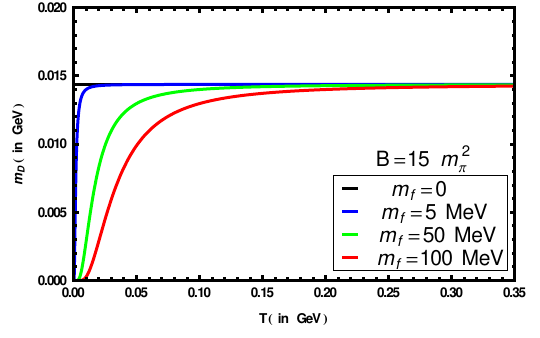}
\hspace{1cm}\includegraphics[scale=0.8]{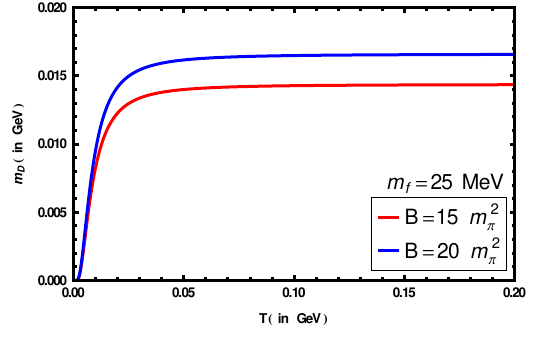}
\end{center}
\caption{{\it Left panel:} Variation of the Debye screening mass with temperature for 
different quark masses massive at a fixed value of $B$.
{\it Right panel:} Comparison of the temperature Variation of the Debye screening mass for 
two values of $B~(=15m_{\pi}^2$ and $20m_{\pi}^2)$.}
\label{ds_plot_1}
\end{figure}

Before discussing the Debye screening we, 
first, note that the effective dimensional reduction in presence of strong magnetic field 
also plays an important role in catalyzing the spontaneous chiral symmetry breaking since 
the fermion pairing takes place in LLL that strengthen the formation of spin-zero 
fermion-antifermion condensates. This enhances the generation of dynamical fermionic mass 
through the chiral condensate in strong field limit even at the weakest attractive 
interaction between fermions~\cite{Shovkovy,gusynin} at $T=0$. 
The pairing dynamics is essentially (1+1) dimensional where the fermion pairs fluctuate in 
the direction of magnetic field. So, the zero temperature magnetized medium is associated 
with 
two scales: the dynamical mass~\footnote{As discussed before we still  represent the  
dynamical mass scale by $m_f$.} $m_f$ and the magnetic field $B$ whereas a hot magnetized 
medium is associated with 
three scales: the dynamical mass $m_f$, temperature $T$ and the magnetic field $B$.

In the left panel of Fig.~\ref{ds_plot_1} the temperature variation of the Debye 
screening mass for quasiquarks in strongly magnetized medium with $B~=15m_{\pi}^2$ and 
for different quark masses is shown. When the quark mass, $m_f=0$, it is found to have a 
finite 
amount of Debye screening. This screening is independent of $T$ because the only scale in 
the system is  the magnetic field ($q_fB~\gg~T^2$), and the thermal scale gets canceled 
out exactly as found analytically  in (\ref{dsmassless}) in contrast to 
Ref.~\cite{alexandre} where one needs to explicitly set the $T\rightarrow 0$ 
limit there. We would like to note that when $T$ drops below the phase 
transition temperature ($T_c$) the screening mass should, in principle,  drop.
However, it is found to remain constant in the region $0 \le T\le T_c$, because of 
the absence of any mass scale in the system.

For massive quarks, the three scales became very distinct and an interesting behavior 
of the Debye screening mass is observed in presence of strong magnetic field. For a 
given $m_f$, as the temperature is being lowered gradually than the value of the fermion 
mass $(T<m_f)$, the quasiquark mass brings the Debye screening down as shown in the left 
panel of Fig.~\ref{ds_plot_1}. Eventually the screening mass vanishes completely 
when $T=0$. When $T\sim m_f$, there is a shoulder in the Debye screening and
as soon as the temperature becomes higher than the value of $m_f$ 
the screening becomes independent of other two scales ($m_f^2\le T^2\le q_f B$). So, in 
presence of strong 
magnetic field the Debye screening mass changes with temperature as long as $T<m_f$ and 
then saturates to a value determined by the strength of the magnetic field.
Further as the quasiquark mass is increased the shoulder and the saturation point are 
pushed towards the higher $T$. The point at which the saturation takes place  depends,
particularly, on the strength of two scales, {\it viz.}, $m_f$ and $T$ associated with 
the 
hot magnetized system.
In other words the dynamical mass generation catalyzes the spontaneous chiral symmetry 
breaking  indicating magnetic catalysis~\cite{Shovkovy,gusynin,alexandre} and in that 
case $T_c$ will be enhanced as a reflection of the dimensionally reduced 
system in presence of strong magnetic field. Now we also note that if the thermal scale is 
higher than the magnetic scale ($T^2~ \gg ~q_fB$), then the Debye screening will increase
with $T$ like the usual hot but unmagnetized medium. For 
this, however, one needs to employ a weak field approximation where higher LL  
contributions will lead to a almost continuous system. This is because in a weak field 
approximation ($q_fB << T^2$), the energy spacing between consecutive Landau levels, 
$[2(n+1) + 1]q_fB - [2n + 1]q_fB = 2q_fB$, gradually reduces with higher levels as shown 
in Fig.~\ref{landau_levels}.
In the right panel a comparison of the Debye screening mass is being shown for massive 
quarks for two values of the magnetic field strength ($B=15m_\pi^2$ and $20m_\pi^2$ ) and 
the screening is enhanced as it is proportional to $B$.


\section{Conclusion and Outlook}
\label{conclu}

In this paper we have evaluated the in-medium electromagnetic spectral function by 
computing  the imaginary part of the photon polarization tensor, in presence of a 
magnetic field. We particularly dealt with the limiting case, where
the magnetic field  is to be very strong  with  respect to the thermal scale 
($q_fB \gg T^2$) of the system. In this strong field limit we have 
exploited the LLL dynamics that decouples the transverse and the longitudinal direction  
as a consequence of an effective dimensional reduction from 
(3+1)-dimension to (1+1)-dimension. 
The electromagnetic spectral function vanishes in the massless limit of quarks  which implies that 
in ($1+1$) dimension an on-shell massless thermal fermion cannot scatter in the forward direction. 
Since the LLL dynamics is (1+1) dimensional, the fermions are virtually paired up in LLL providing a strongly 
correlated system, which could possibly enhance the generation of fermionic mass through the chiral condensate.
So, these massive quarks could provide a kinematical threshold to the electromagnetic spectral function at longitudinal photon momentum, 
$p_\shortparallel^2=4m_f^2$. Below the threshold the photon polarization tensor is purely real and the electromagnetic 
spectral function does not exist resulting in no pair creation of particle and antiparticle. 
This implies that the momentum of the external photon supplies energy to virtual 
fermionic pairs in LLL, which become real via photon decay. At threshold the photon strikes the LLL and 
the spectral strength diverges due to the dimensional reduction, since a factor of $\left(1-{4m_f^2}/{p_\shortparallel^2}\right)^{-{1}/{2}}$
appears in the spectral function, in strong field approximation. The spectral strength starts with a high value for the photon longitudinal 
momentum $p_\shortparallel > 2m_f$  due to the dimensional reduction or LLL dynamics and then falls off with increase of $\omega$ as there 
is nothing beyond the LLL in strong field approximation. 

This strong field approximation could possibly be very appropriate for the initial stages 
of the noncentral heavy-ion collisions where the intensity of the produced magnetic field 
is expected to be very high
As a spectral property we then obtained analytically  the 
dilepton production rate for two scenarios: (i) the quarks and antiquarks are affected 
by the hot magnetized medium but not the final lepton pairs and (ii) when both  quark 
and lepton are affected by the magnetized medium.
In the former case the dilepton rate is ${\cal O}[|q_fB|]$ and follows the 
properties of the electromagnetic spectral function along with a kinematical threshold 
provided by the quark mass. For the later case the rate is found to be ${\cal O}[|eB|^2$ 
with two kinematical thresholds provided by quark ($m_f$) and lepton ($m_l$) mass. Since 
the dynamics in LLL in strong filed approximation is strongly correlated one, the 
threshold will finally be determined by $\tilde m= {\mbox{max}}(m_f,m_l)$. 

We have also analyzed the electromagnetic screening effect through  the Debye screening  
mass of the hot magnetized medium. This shows that there are three distinct scales in a 
hot magnetized medium,  associated with the mass of the quasiquarks,  
temperature of the medium and the background magnetic field strength. 
When the mass of the 
quasiquarks are much higher than the temperature, the Debye screening is negligible. As 
the temperature increases, the screening mass starts increasing, a shoulder like 
structure appears when $T\sim m_f$,  and then it saturates to a fixed value when $q_fB \gg 
T^2\gg m_f^2$. In a strongly magnetized hot medium the Debye screening mass shows an 
interesting 
characteristics with temperature as long as $T\le m_f$ and 
then saturates to a value determined by the strength of the magnetic field.
The point at which the saturation takes place  depends, especially,  on 
the strength  of mass and temperature  scale associated with a hot 
magnetized system.
In strong field approximation  the fermion pairing takes place in LLL 
that could enhance the formation of  
quark-antiquark condensates, leading to a larger dynamical mass generation which 
catalyzes the spontaneous chiral symmetry breaking.
This mass effect is reflected in the Debye screening as
the shoulder and the saturation point are pushed towards a higher $T$  when the 
quasiquark mass increases. The effective dimensional reduction 
seems to plays an important role in catalyzing the spontaneous chiral symmetry breaking, 
which indicates an occurrence of magnetic catalysis effect in presence of strong 
magnetic field.

\section{Acknowledgements}

This work is supported by the Department of Atomic Energy (DAE), India through the 
project TPAES. Authors acknowledge fruitful discussion with A. Ayala and M. Strickland. 
AB gratefully acknowledges useful discussions with P. Chakraborty.

\appendix
\renewcommand{\theequation}{\thesection.\arabic{equation}}

\section{Processes with (a) $r=1,l=-1$  and (b) $r=-1,l=1$ }
\label{app_a}
So, choosing first  $r=1,l=-1$ we obtain from (\ref{Pi_sfa_gen})

\bea
\mathcal{I}m~\Pi_\mu^\mu(\omega,{\bf p})\Big\vert_{r=1\atop s=-1}\!\! \!\! &=& \! \! 
N_c\pi \sum_{f}e^{\frac{-p_\perp^2}{2q_fB}}~~\frac{2q_f^3Bm_f^2}{\pi}\int\frac{dk_3}{2\pi}
\frac{\left(1-n_F(E_k)\right)\left(1-n_F(-E_q)\right)}{4E_kE_q} \nn\\
&& \times \left[e^{ -\beta(E_k-E_q)}-1\right] \delta(p_0-E_k+E_q). \label{a1}
\eea

Now, using $1-n_F(-E_q)=n_F(E_q)$, one obtains
\bea
 \mathcal{I}m~\Pi_\mu^\mu(\omega,{\bf p})\Big\vert_{r=1\atop s=-1} &=& 
N_c\sum_{f}
e^{\frac{-p_\perp^2}{2q_fB}}~~\frac{2q_f^3Bm_f^2}{\pi}
\int\frac{dk_3}{2}~\delta(\omega-E_k+E_q)\frac{\left[n_F(E_k)-n_F(E_q)\right]}{4E_kE_q}.
\label{a2}
\eea

The $k_3$ integral can now be performed  using the following property of the delta 
function 
\bea
\int\limits_{-\infty}^{\infty} dp_3~ f(p_3)~ \delta[g(p_3)] = \sum_{r} 
\frac{f(p_{zr})}{\vert g^\prime(p_{zr})\vert}, \label{a3}
\label{delta_property}
\eea
where the zeroes of the argument inside the delta function is called as $p_{zr}$.

Now $\omega-E_k+E_q = 0 $ yields,
\bea
k_3^z = \frac{p_3}{2} \pm \frac{\omega}{2}\sqrt{1-\frac{4m_f^2}{(\omega^2-p_3^2)}},
&=& \frac{p_3}{2} \pm \frac{\omega R}{2}, \label{a4}\\
\vert g^\prime(p_{z})\vert &=& 
\Big\vert\frac{E_k(k_3-p_3)-E_qk_3}{E_kE_q}\Big\vert_{k_3=k_3^{z1},k_3^{z2}}~~, 
\label{a5} \\
E_k\Big\vert_{k_3=k_3^{z1}} = \frac{\omega}{2} + \frac{p_3 R}{2}; && 
E_k\Big\vert_{k_3=k_3^{z2}} = \frac{\omega}{2} - \frac{p_3 R}{2},\label{a6} \\
E_q\Big\vert_{k_3=k_3^{z1}} = \frac{\omega}{2} - \frac{p_3 R}{2}; && 
E_q\Big\vert_{k_3=k_3^{z2}}=\frac{\omega}{2} + \frac{p_3 R}{2},\label{a7} \\
\mbox{and} \, \, \Big\vert E_k(k_3-p_3)-E_qk_3\Big\vert_{k_3=k_3^{z1},k_3=k_3^{z2}} &=& 
\frac{\omega p_3}{2}(R^2-1).\label{a8}
\eea

\bea
 \mathcal{I}m~\Pi_\mu^\mu(\omega,{\bf p})\Big\vert_{r=1\atop s=-1} &=& N_c\sum_{f}
e^{\frac{-p_\perp^2}{2q_fB}}~~\frac{2q_f^3Bm_f^2}{\pi}\sum\limits_r\frac{\left[
n_F(E_k)-n_F(E_q)\right]}{8E_kE_q}\times 
\Big\vert\frac{E_kE_q}{E_k(k_3-p_3)-E_qk_3}\Big\vert\Bigg\vert_{k_3=k_3^{zr}}~~ \nn\\
&=& N_c\sum_{f}
e^{\frac{-p_\perp^2}{2q_fB}}~~\frac{2q_f^3Bm_f^2}{\pi}\sum\limits_r\frac{\left[
n_F(E_k)-n_F(E_q)\right]}{8\vert E_k(k_3-p_3)-E_qk_3\vert}\Bigg\vert_{k_3=k_3^{zr}}~~ 
\nn\\
&=& N_c\sum_{f}
e^{\frac{-p_\perp^2}{2q_fB}}~~\frac{2q_f^3Bm_f^2}{4\pi\omega p_3(R^2-1)}\times\nn\\
&&\left[n_F(E_k\Big\vert_{k_3=k_3^{z1}})-n_F(E_q\Big\vert_{k_3=k_3^{z1}}
)+n_F(E_k\Big\vert_{k_3=k_3^{z2}})-n_F(E_q\Big\vert_{k_3=k_3^{z2}})\right]\nn\\
&=& N_c\sum_{f}
e^{\frac{-p_\perp^2}{2q_fB}}~~\frac{2q_f^3Bm_f^2}{4\pi\omega p_3(R^2-1)}\times \nn\\
&&\left[n_F\left(\frac{\omega}{2} + \frac{p_3 R}{2}\right)-n_F\left(\frac{\omega}{2} - 
\frac{p_3 R}{2}\right)+n_F\left(\frac{\omega}{2} - \frac{p_3 
R}{2}\right)-n_F\left(\frac{\omega}{2} + \frac{p_3 R}{2}\right)\right] \nn\\
&=& 0. \label{a9}
\eea
Similarly, for the case (b) $r=-1,l=1$, the phase space also does not 
allow the corresponding process. 




\begin{thebibliography}{99}

\bibitem{Shovkovy}
I. A. Shovkovy,
Lect. Notes Phys. {\bf 871}, 13 (2013), 
arXiv:1207.5081 [hep-ph]. 

\bibitem{Elia}
M. D’Elia,
Lect. Notes Phys. {\bf 871}, 181 (2013), 
arXiv:1209.0374 [hep-lat]. 

\bibitem{Fukushima}
K. Fukushima, 
Lect. Notes Phys. {\bf 871}, 241 (2013), 
arXiv:1209.5064 [hep-ph]. 

\bibitem{Mueller}
N. Mueller, J. A. Bonnet, and C. S. Fischer, 
Phys.\ Rev.\ D {\bf 89}, 094023 (2014), 
arXiv:1401.1647 [hep-ph]. 

\bibitem{Miransky}
V. A. Miransky and I. A. Shovkovy, 
Physics Reports {\bf 576},1-209 (2015), 
arXiv:1503.00732 [hep-ph].

\bibitem{bzdak}
A.~Bzdak and V.~Skokov,
Phys.\ Rev.\ Lett. {\bf 110}, 192301 (2013),
arXiv:1208.5502 [hep-ph].

\bibitem{McLerran}
L.~McLerran and V.~Skokov
Nucl.\ Phys.\ A {\bf 929}, 184, (2014),
arXiv:1305.0774 [hep-ph].

\bibitem{Tuchin:2013prc2}
  K.~Tuchin,
  Phys.\ Rev.\ C {\bf 87}, 024912 (2013).

 \bibitem{Tuchin:2013prc}
  K.~Tuchin,
  Phys.\ Rev.\ C {\bf 88}, 024910 (2013),
  arXiv:1305.0545 [nucl-th].
 
\bibitem{Tuchin:2013ahep} K.~Tuchin, Adv.High Energy Phys. 2013 (2013).  
   
\bibitem{cme1} 
 D.~E.~Kharzeev, L.~D.~McLerran and H.~J.~Warringa,
 Nucl.\ Phys.\ A {\bf 803}, 227 (2008),
 arXiv:0711.0950 [hep-ph].
 
 \bibitem{cme2} 
 K.~Fukushima, D.~E.~Kharzeev and H.~J.~Warringa,
 Phys.\ Rev.\ D {\bf 78}, 074033 (2008),
 arXiv:0808.3382 [hep-ph].
 
 \bibitem{cme3} 
 D.~E.~Kharzeev,
 Ann.\ Phys. {\bf 325}, 205 (2010),
 arXiv:0911.3715 [hep-ph].
 
 \bibitem{mcat1}
 J.~Alexandre, K.~Farakos and G.~Koutsoumbas,
 Phys.\ Rev.\ D {\bf 63}, 065015 (2001),
 hep-th/0010211. 

 \bibitem{mcat2}
 V.~P.~Gusynin and I.~A.~Shovkovy, 
 Phys.\ Rev.\ D {\bf 56}, 5251 (1997),
 hep-ph/9704394.

 \bibitem{mcat3}
 D.~S.~Lee, C.~N.~Leung, and Y.~J.~Ng, 
 Phys.\ Rev.\ D {\bf 55}, 6504 (1997),
 hep-th/9701172.

\bibitem{Bali} 
G.~S.~Bali, F.~Bruckmann, G.~Endrodi, Z.~Fodor, S.~D.~Katz, S.~Krieg, A.~Schafer and K.~K.~Szabo,
JHEP {\bf 1202}, 044 (2012),
 arXiv:1111.4956 [hep-lat].

\bibitem{Bornyakov}
V.~G.~Bornyakov, P.~V.~Buividovich, N.~Cundy, O.~A.~Kochetkov and A.~Schafer,
Phys. Rev. D {\bf 90}, 034501 (2014),
arXiv:1312.5628 [hep-lat]
 
 \bibitem{mueller}
 N.~Mueller and J.~M.~Pawlowski,
 arXiv:1502.08011v2 [hep-ph].
 
\bibitem{Ayala:2014iba} 
A. Ayala, M. Loewe,  A. Z. Mizher and Zamora, R.,
Phys. Rev.D {\bf 90}, 036001  (2014)
 
 \bibitem{Ayala:2014gwa} 
  A.~Ayala, M.~Loewe and R.~Zamora,
  Phys.\ Rev.\ D {\bf 91}, 016002 (2015)
  [arXiv:1406.7408 [hep-ph]].
 
 \bibitem{Ayala:2016sln} 
  A.~Ayala, M.~Loewe and R.~Zamora,
  J.\ Phys.\ Conf.\ Ser.\  {\bf 720}, 012026 (2016).
  
  \bibitem{Ayala:2015bgv} 
  A.~Ayala, C.~A.~Dominguez, L.~A.~Hernandez, M.~Loewe and R.~Zamora,
  Phys.\ Lett.\ B {\bf 759}, 99 (2016),
  [arXiv:1510.09134 [hep-ph]].
 
 \bibitem{fayazbakhsh1}
 S.~Fayazbakhsh and N.~Sadooghi,
 Phys. Rev. D. {\bf 83}, 025026 (2011),
 arxiv:1009.6125 [hep-ph].
 
 \bibitem{fayazbakhsh2}
 S.~Fayazbakhsh and N.~Sadooghi,
 Phys. Rev. D. {\bf 82}, 045010 (2010),
 arxiv:1005.5022 [hep-ph].
 
 \bibitem{jens1}
 J.~O.~Andersen, W.~R.~Naylor and A.~Tranberg,
 arxiv:1411.1176 [hep-ph].
 
 \bibitem{mike1}
 M.~Strickland, V.~Dexheimer and D.~P.~Menezes,
 Phys. Rev. D. {\bf 86}, 125032 (2012),
 arxiv:1209.3276 [hep-ph].
 
 \bibitem{jens_rmp}
 J.~O.~Andersen, W.~R.~Naylor and A.~Tranberg,
 Rev. Mod. Phys. {\bf 88}, 025001 (2016),
 arXiv:1411.7176 [hep-ph].
 
 \bibitem{fayazbakhsh3}
 S.~Fayazbakhsh, S.~Sadeghian and N.~Sadooghi,
 Phys. Rev. D. {\bf 86}, 085042 (2012),
 arxiv:1206.6051 [hep-ph].
 
 \bibitem{fayazbakhsh4}
 S.~Fayazbakhsh and N.~Sadooghi,
 Phys. Rev. D. {\bf 88}, 065030 (2013),
 arxiv:1306.2098 [hep-ph].
 
  \bibitem{basar}
 G.~ Basar, D.~ E.~ Kharzeev, and V.~ Skokov,
 Phys.\ Rev.\ Lett. {\bf 109}, 202303 (2012),
 arXiv:1206.1334 [hep-ph].
 
 \bibitem{Ayala:2016lvs} 
  A.~Ayala, J.~D.~Castano-Yepes, C.~A.~Dominguez and L.~A.~Hernandez,
  arXiv:1604.02713 [hep-ph].
 
 \bibitem{sadooghi1} 
 N. Sadooghi and F. Taghinavaz, 
 Phys. Rev. D. {\bf 92}, 025006 (2015),
 arXiv:1504.04268 [hep-ph].
 
 \bibitem{sadooghi} N. Sadooghi and F. Taghinavaz, 
 arXiv:1601.04887 [hep-ph].
 
  \bibitem{mamo}
 K. A. Mamo, JHEP {\bf 1308}, 083 (2013).
 
 \bibitem{phenix}
 A.~ Adare et al. (PHENIX Collaboration),
 Phys.\ Rev.\ Lett. {\bf 109}, 122302 (2012),
 arXiv:1105.4126 [nucl-ex].
 
 \bibitem{rojas}
 H.~P.~Rojas and A.~E.~Shabad,
 Ann.\ Of.\ Phys. {\bf 138}, 1-35 (1982).
 
 \bibitem{hattori}
 K. Hattori  and  K. Itakura,
 Ann. Phys.  {\bf 330}, 23 (2013); {\it ibid.} {\bf 334}, 58 (2013),
 arXiv:1212.1897 [hep-ph].
 
 \bibitem{chao}
 J.~Chao, L.~Yu and M.~Huang,
 Phys.\ Rev.\ D {\bf 90},  045033 (2014),
 arXiv:1403.0442 [hep-th].

 \bibitem{Tsai1}
Wu-yang Tsai,
Phys.\ Rev.\ D {\bf  10}, 2699 (1974).

\bibitem{Tsai2}
Wu-yang Tsai and T.~Erber,
Phys.\ Rev.\ D {\bf  10}, 492 (1974).

\bibitem{jackson} J. Jackson, \textit {Classical Electrodynamics}, 
2nd ed. ~Wiley, New
York, 1975; R. Dalitz and D. Yennie, Phys. Rev. {\bf 105}, 1598  (1957).

\bibitem{ritus}
V. I. Ritus, Ann. phys. {\bf 69}, 555 (1972).
 
 \bibitem{schwinger}
 J.~Schwinger,
 Phys.\ Rev.\ {\bf 82}, 664 (1951).

  \bibitem{gusynin} 
 V.~P.~Gusynin, V.~A.~Miransky and I.~A.~Shovkovy,
 Nucl.\ Phys.\ B {\bf 462}, 249 (1996),
 hep-ph/9509320.
 
 \bibitem{calucci} 
 G.~Calucci and R.~Ragazzon,
 J.\ Phys.\ A; Math.\ Gen.\ {\bf 27}, 2161 (1994).  
 
\bibitem{ferrer} E. J. Ferrer and V. de la Incera, 
Phys. Rev {\bf D 58}, 065008 (1998).

\bibitem{peskin} M. E. Peskin and D. V. Schroeder, 
{\it An Introduction to Quantum Field Theory}, Westview Press, 1995.

\bibitem{aminul}
Ch. A. Islam, S. Majumder, N. Haque and M. G. Mustafa,
JHEP 1502 (2015) 011.
 
\bibitem{najmul}
C. Greiner, N. Haque, M. G. Mustafa and M. H. Thoma,
Phys. Rev. {\bf C83},  014908 (2011).
 
 \bibitem{pisarski}
 R.~D.~Pisarski,
 Nucl.\ Phys.\ B {\bf 309}, 476 (1988).
 
 \bibitem{Das:2012pd} 
  A.~Das, R.~R.~Francisco and J.~Frenkel,
  Phys.\ Rev.\ D {\bf 86}, 047702 (2012),
  arXiv:1206.5677 [hep-th].
  
\bibitem{Weldon:1990iw} H. A. Weldon, Phys. Rev. D{\bf 42}, 2348 (1990).

\bibitem{landau} L. D. Landau and Lifshitz, \textit{Quantum Mechanics: Non-
relativistic Theory (Pergamon Press, 1977)}.


\bibitem{kapusta} J.I. Kapusta, \textit{Finite-temperature Field Theory} 
(Cambridge University Press, Cambridge, 1989).

\bibitem{lebellack} M. Le Bellac, 
\textit{Thermal Field Theory} (Cambridge University Press, Cambridge, 2000).

\bibitem{alexandre}
 J.~Alexandre,
 Phys.\ Rev.\ D {\bf 63}, 073010 (2001),
 hep-th/0009204.
 
\end{thebibliography}
\end{document}